\documentclass[aps,pra,twocolumn,showpacs,preprintnumbers,amsmath,amssymb]{revtex4-1}
\usepackage{graphicx}
\usepackage[usenames,dvipsnames]{xcolor}
\usepackage{cleveref}
\usepackage{makecell}
\usepackage{float}
\usepackage{lineno}
\usepackage[caption=false]{subfig}
\newcommand{\bra}[1]{\ensuremath{\langle #1 |}}
\newcommand{\ket}[1]{\ensuremath{| #1 \rangle}}

\begin{document}
%\preprint{DRAFT}
\title{Truncation effects in the charge representation of the O(2) model}

\author{Jin Zhang$^{1}$}
\email{jin-zhang@uiowa.edu}
\author{Y. Meurice$^1$}
\author{S.-W. Tsai$^2$}
\affiliation{$^1$Department of Physics and Astronomy, University of Iowa, Iowa City, IA 52242, USA}
\affiliation{$^2$Department of Physics and Astronomy, University of California, Riverside, CA 92521, USA}
\definecolor{burnt}{cmyk}{0.2,0.8,1,0}
\def\lt{\lambda ^t}
\def\note{note}
\def\beq{\begin{equation}}
\def\enq{\end{equation}}

\date{\today}
\begin{abstract}
The O(2) model in Euclidean space-time is the zero-gauge-coupling limit of the compact scalar quantum electrodynamics. We obtain a dual representation of it called the charge representation. We study the quantum phase transition in the charge representation with a truncation to ``spin $S$," where the quantum numbers have an absolute value less than or equal to $S$. The charge representation preserves the gapless-to-gapped phase transition even for the smallest spin truncation $S = 1$. The phase transition for $S = 1$ is an infinite-order Gaussian transition with the same critical exponents $\delta$ and $\eta$ as the Berezinskii-Kosterlitz-Thouless (BKT) transition, while there are true BKT transitions for $S \ge 2$. The essential singularity in the correlation length for $S = 1$ is different from that for $S \ge 2$. The exponential convergence of the phase-transition point is studied in both Lagrangian and Hamiltonian formulations. We discuss the effects of replacing the truncated $\hat{U}^{\pm} = \exp(\pm i \hat{\theta})$ operators by the spin ladder operators $\hat{S}^{\pm}$ in the Hamiltonian. The marginal operators vanish at the Gaussian transition point for $S = 1$, which allows us to extract the $\eta$ exponent with high accuracy.
\end{abstract}

\maketitle

%%%%%%%%%%%%%%%%%%%%%%%%%%%%%%%%%%%%%%%%%%%%%%%%%%%%%%%%%%%%%%%%%%%%%%%%%%
\section{Introduction}\label{sec:introduction}
%%%%%%%%%%%%%%%%%%%%%%%%%%%%%%%%%%%%%%%%%%%%%%%%%%%%%%%%%%%%%%%%%%%%%%%%%%
There has been remarkable recent progress \cite{Weimer2010, Cirac2012, Bloch2012, Bernien2017, Monroe2021RMP, PRXQuantum.2.017003, Banuls2020} in the development of controllable quantum systems into quantum simulators to recreate and test model Hamiltonians, and, importantly, to provide answers to open questions that cannot be solved with classical computers. These efforts may help elucidate properties of complex quantum materials, questions involving the dynamics and quantum critical phenomena, as well as problems relevant to nuclear and high-energy physics. Lattice gauge theory (LGT) offers interesting models that are introduced in high-energy physics as cutoff-regularized formulations of gauge theories used to describe strongly interacting particles. Mappings of gauge-field theories into lattice Hamiltonians of particles or spins \cite{RevModPhys.51.659} then allow for the possibility of quantum simulating these LGT models\cite{PhysRevD.92.076003,PhysRevLett.121.223201,PhysRevD.98.094511} in the laboratory. For models with continuous symmetry, the mapping leads to discrete quantum numbers for the effective Hamiltonian that needs to be truncated for quantum simulation. The truncations themselves may correspond to a series of interesting models such as clock models \cite{PhysRevB.98.205118, PhysRevB.100.094428}, spin-$S$ models \cite{PhysRevLett.108.127202}, and boson models \cite{Dutta_2015}. A key question that arises is how the truncation affects the critical properties of the model for a given formulation of the mapping. In this paper, we address this question for the O(2) model by investigating the critical behavior of its dual representation, the effects of truncation and implications for its quantum simulation. We find that spin truncations as small as $S=2$ already captures its correct critical behavior. The $S=1$ truncation exhibits a multicritical point corresponding to an infinite-order Gaussian transition that, while not capturing the behavior of the O(2) model, is interesting in its own right.

Topological excitations (instantons, monopoles, and vortices) play an important role in the physics of gauge theories with continuous symmetry groups. The classical planar model in two dimensions or the O(2) model in $(1+1)$-dimensional Euclidean space-time, was the first discovered to have a Berezinskii-Kosterlitz-Thouless (BKT) transition \cite{Berezinsky:1970fr, Kosterlitz_1973, Kosterlitz_1974}. The BKT transition is a very important infinite-order phase transition in two-dimensional systems with continuous symmetry. In 1974, Kosterlitz performed a renormalization group (RG) analysis of the O(2) model and found an essential singularity in the correlation length and the same critical exponents $\delta, \eta$ as the two-dimensional Ising universality class \cite{Kosterlitz_1974}. In one-dimensional quantum systems, the equivalence of path integral quantization to statistical mechanics in two dimensions assures that the same type of phase transitions can happen driven by quantum fluctuations. In ($2+1$) dimensions, various exotic phenomena in both condensed-matter and high-energy physics belongs to the BKT universality class, including the superfluid transitions in two-dimensional ($2$D) Bose gases \cite{Hung2011, PhysRevLett.110.145302}, superconducting transitions in $2$D materials \cite{PhysRevLett.107.217003, Sharma2018, PhysRevLett.123.075303}, and the confinement-deconfinement phase transitions in U(1) lattice gauge theories \cite{PhysRevD.53.7157, GRIGNANI1996143, KLEINERT2003361}.

Due to the essential singularity and the logarithmic corrections stemming from the marginal operator, it is difficult to detect the BKT transitions accurately using classical methods. The O(2) model also can be seen as the zero-gauge-coupling limit of compact scalar quantum electrodynamics (sQED) and there have been proposals to quantum-simulate it \cite{PhysRevA.90.063603, PhysRevD.92.076003, PhysRevA.96.023603}. Because quantum simulators have discrete variables, proposals transform the O(2) model into the discrete dual space, where the discrete variables have the physical meaning of charge and current quantum numbers \cite{PhysRevD.88.056005,PhysRevE.89.013308}. By applying Gauss's law, we can also go into the discrete space expanded by electric-field quantum numbers \cite{PhysRevD.98.094511, PhysRevLett.121.223201}. These quantum numbers take integers from $-\infty$ to $+\infty$, so a truncation is needed for quantum simulations. Some truncation effects in the mass gap and $\beta$ functions for the O(2) model have been discussed in Ref. \cite{PhysRevD.93.085012}. In this paper, we discuss how truncation affects the quantum phase transition in the charge representation of the O(2) model in detail. We are interested in determining the most economical truncation that allows us to probe the finite-size effect of the original O(2) model and preserves the BKT transition. We investigate how the phase transitions change with the size of the truncation. Truncation effects in a sequence of models, such as the O(3) model with a chemical potential \cite{PhysRevD.99.074501}, the O(4) model \cite{PhysRevD.93.085012}, the Schwinger model \cite{PhysRevD.95.094509}, and the SU(2) lattice gauge theory \cite{PhysRevX.7.041046}, have been studied in the context of lattice QCD, but the truncation effects on the BKT transition in the O(2) model has not been studied in detail in previous works. Our work employs a combination of numerical approaches to determine the truncation effects, and fills this gap in the progress towards quantum simulation of lattice QCD \cite{meurice2020tensor}.

Efficient tools to detect BKT transitions are needed to study these truncation effects. Level spectroscopy (LS) is one of the most efficient and accurate methods for systems with a small dimension of the local Hilbert space \cite{Nomura_1995, PhysRevLett.76.4038, doi:10.1143/JPSJ.66.3944, doi:10.1143/JPSJ.66.3379, Nomura_1998}. Accurate phase diagrams for various models have been determined by LS more than $15$ years ago even with modest computational resources \cite{PhysRevLett.76.4038, doi:10.1143/JPSJ.66.3944, doi:10.1143/JPSJ.66.3379, PhysRevB.67.104401}. The LS technique requires a detailed analysis of the scaling dimensions of different types of excitations near the BKT critical line and the determination of which levels cross. The level crossing can be between different types of excitations for different systems. After obtaining the results from LS, we test a universal method to detect quantum phase transitions: the scaling of the energy gap between the ground state and the first-excited state \cite{PhysRevB.84.115135, PhysRevA.87.043606, PhysRevB.91.165136, PhysRevB.99.134408, PhysRevB.102.064414}. The truncation effects in this energy gap that contains information on the divergent behavior of the correlation length, are also important indicators for different types of phase transitions. The method does not require a prior knowledge of the critical properties for the target systems and only requires a bulk quantity: the energy gap between the lowest two levels that can be obtained accurately by the density-matrix-renormalization-group (DMRG) algorithm \cite{PhysRevLett.69.2863, PhysRevB.48.10345, RevModPhys.77.259, SCHOLLWOCK201196}. Various works \cite{PhysRevB.16.1153, PhysRevB.23.4698, Jullien_1981, DENNIJS1982273, PhysRevB.30.1629, doi:10.1063/1.333675, PIRES2007387} have shown similarities and differences between the phase transition in $S = 1$ and BKT transitions, with some disagreement in the location of the phase transition and the form of the critical scaling of the correlation length. We give concrete evidence for the difference in the essential singularity between $S = 1$ and $S \ge 2$ and accurate infinite-order phase-transition points by performing LS and DMRG calculations.

The paper is organized as follows: Section~\ref{subsec:o2model} introduces the O(2) model and the origin of the charge representation. The Hamiltonian in the charge representation and its properties are described in Sec. \ref{subsec:chargerepre}. We compare the truncated $\hat{U}^{\pm}$ operators and the spin ladder operators $\hat{S}^{\pm}$ in Sec. \ref{subsec:upmandspm}. In Sec. \ref{subsec:modells}, we introduce the LS technique. The ansatz for the scaling of the energy gap as a universal tool to detect quantum phase transitions is introduced in Sec. \ref{subsec:modelgs}. The parameters used in numerical algorithms are given in Sec. \ref{subsec:modeltrgdmrg}. We discuss the results in Sec. \ref{sec:results}. Section~\ref{subsec:lagmagneticsus} presents the results from the Lagrangian. The determination of phase-transition points with LS for the Hamiltonian is discussed in Sec. \ref{subsec:hamls}. In Sec. \ref{subsec:resultsgaps}, we use the ansatz of the scaling of the energy gap to locate the phase-transition points and compare them with those from LS. We emphasize the difference between $S = 1$ and $S \ge 2$ in each part of Sec. \ref{sec:results}. Finally, in Sec. \ref{sec:conclusion}, we summarize the main conclusions of our work and point out possible future work.

%%%%%%%%%%%%%%%%%%%%%%%%%%%%%%%%%%%%%%%%%%%%%%%%%%%%%%%%%%%%%%%%%%%%%%%%%%
\section{Model and Methods}\label{sec:model}
%%%%%%%%%%%%%%%%%%%%%%%%%%%%%%%%%%%%%%%%%%%%%%%%%%%%%%%%%%%%%%%%%%%%%%%%%%

%%%%%%%%%%%%%%%%%%%%%%%%%%%%%%%%%%%%%%%%%%%%%%%%%%%%%
\subsection{Action} \label{subsec:o2model}
%%%%%%%%%%%%%%%%%%%%%%%%%%%%%%%%%%%%%%%%%%%%%%%%%%%%%

On a $(L-1) \times L_\tau$ Euclidean lattice, the action of the O(2) model is
\begin{eqnarray}
\label{eq:o2action}
S = - \sum_{\mu = \tau, s} \beta_\mu \sum_{\mathbf{x}} \cos(\theta_{\mathbf{x}+\hat{\mu}} - \theta_{\mathbf{x}}) - h \sum_{\mathbf{x}} \cos(\theta_{\mathbf{x}}),
\end{eqnarray}
where $\beta_{\tau(s)}$ is a coupling constant in the temporal (spatial) direction, $\mathbf{x} = \left(x_s, x_\tau \right)$ is the 2D position vector, and $\hat{\tau}$ ($\hat{s}$) is the unit vector in the temporal (spatial) direction. In the isotropic case, $\beta_\tau = \beta_s = \beta$ is the inverse temperature $1/T$ in the context of statistical mechanics. The parameter $h$ is an external field. The path integral formulation is written as
\begin{eqnarray}
\label{eq:pathintegral}
Z = \int \prod_{\mathbf{x}} \frac{d \theta_{\mathbf{x}}}{2\pi} e^{-S}.
\end{eqnarray}
By expanding the weights with modified Bessel functions, Eq. \eqref{eq:pathintegral} can be rewritten as \cite{RevModPhys.52.453, PhysRevD.88.056005}
\begin{eqnarray}
\label{eq:o2pathintegralbessel}
\nonumber Z = && I_0(\beta)^{2V} I_0(h)^V \sum_{l_{\mathbf{x}} = n_{\mathbf{x},s\tau}} \prod_{\mathbf{x}} t_{n_{\mathbf{x},\tau}}(\beta_\tau) t_{n_{\mathbf{x},s}}(\beta_s) t_{l_{\mathbf{x}}}(h) \\ \propto && \sum_{n_{\mathbf{x},\tau}, n_{\mathbf{x},s}} \prod_{\mathbf{x}} \mathcal{A}_{n_{\mathbf{x}-\hat{\tau},\tau}, n_{\mathbf{x}-\hat{s},s}, n_{\mathbf{x},\tau}, n_{\mathbf{x},s}},
\end{eqnarray}
where the volume $V = (L-1) L_{\tau}$, the summations are over $\{n_{\mathbf{x},\tau}, n_{\mathbf{x},s}\}$ with the condition $l_{\mathbf{x}} = n_{\mathbf{x},s\tau} = n_{\mathbf{x},\tau} + n_{\mathbf{x},s} - n_{\mathbf{x}-\hat{\tau},\tau} - n_{\mathbf{x}-\hat{s},s}$, $t_n(x) = I_n(x) / I_0(x)$, and $I_n(x)$ is the $n$th-order modified Bessel function of the first kind. The four-rank tensor
\begin{eqnarray}
\label{eq:partitionfbessel}
&& \mathcal{A}_{n_{\mathbf{x}-\hat{\tau},\tau}, n_{\mathbf{x}-\hat{s},s}, n_{\mathbf{x},\tau}, n_{\mathbf{x},s}} \nonumber \\
&=& \sqrt{t_{n_{\mathbf{x}-\hat{\tau},\tau}}(\beta_\tau)t_{n_{\mathbf{x}-\hat{s},s}}(\beta_s)t_{n_{\mathbf{x},\tau}}(\beta_\tau)t_{n_{\mathbf{x},s}}(\beta_s)}t_{l_{\mathbf{x}}}(h).
\end{eqnarray}
In the context of sQED, $n_{\mathbf{x},\tau}$ and $n_{\mathbf{x},s}$ have the physical meaning of charge and current quantum numbers, respectively, and are attached to the links of the space-time lattice. We call Eq. \eqref{eq:o2pathintegralbessel} the charge representation of the path integral quantization. Without external field, the sum of $l_{\mathbf{x}}$ with time coordinate fixed at $x_{\tau} = x_0$, $\sum_{\mathbf{x}, x_\tau = x_0} l_{\mathbf{x}} = 0$, giving $\sum_{\mathbf{x},x_\tau=x_0} n_{\mathbf{x},\tau} = \sum_{\mathbf{x},x_\tau=x_0} n_{\mathbf{x}-\hat{\tau},\tau}$. Thus the total charges in any two nearest-time slices at $x_{\tau} = x_0, x_0-1$ are equal and therefore conserved. The charge representation contains all charge sectors for both periodic boundary conditions (PBCs) and open boundary conditions (OBCs). Note that there are $L-1$ plaquettes and $L$ links in a time slice for this configuration. The tensor reformulation of the expectation value of an observable can be obtained in the same way. For example, the magnetization $M = \langle \cos(\theta_{\mathbf{x}^*})\rangle$ is expressed as
\begin{eqnarray}
\label{eq:tensormag}
M = \frac{\sum_{l_{\mathbf{x}^*} = n_{\mathbf{x}^*,s\tau}-1} \sum_{l_{\mathbf{x} \neq \mathbf{x}^*} = n_{\mathbf{x},s\tau}} \prod_{\mathbf{x}} \mathcal{A}}{\sum_{l_{\mathbf{x}} = n_{\mathbf{x},s\tau}} \prod_{\mathbf{x}} \mathcal{A}},
\end{eqnarray}
where $\mathbf{x}^*$ is the position of the local spin $\cos(\theta_{\mathbf{x}^*})$. Thus the tensor contraction in the numerator of Eq. \eqref{eq:tensormag} has an impure tensor at $\mathbf{x}^*$.

%%%%%%%%%%%%%%%%%%%%%%%%%%%%%%%%%%%%%%%%%%%%%%%%%%%%
\subsection{Hamiltonian in the charge representation} \label{subsec:chargerepre}
%%%%%%%%%%%%%%%%%%%%%%%%%%%%%%%%%%%%%%%%%%%%%%%%%%%%
Based on the equivalence of the two-dimensional statistical mechanics and the ($1+1$)-dimensional quantum field theory, we can study the model using the Hamiltonian approach. Following Refs. \cite{PhysRevD.92.076003, PhysRevD.98.094511}, we obtain the Hamiltonian in the charge representation
\begin{align}
	\label{eq:any-spin-ham-charge}
	\hat{H}_{c} = \frac{Y}{2} \sum_{l=1}^{L} (\hat{S}_{l}^z)^2 - \frac{X}{2} \sum_{l = 1}^{L-1} (\hat{U}_{l}^+ \hat{U}_{l+1}^- + \hat{U}_{l}^- \hat{U}_{l+1}^+)
\end{align}
where $Y = 1/\beta_\tau a_\tau, X = \beta_s / a_\tau$, $a_\tau \rightarrow 0$ is the lattice spacing in the temporal direction, and the limit $\beta_\tau \rightarrow \infty, \beta_s \rightarrow 0$ is taken to keep $Y$ and $X$ finite. The index $l$ labels the links. $\hat{S}^z$ is the electric charge operator satisfying $\hat{S}^z \ket{n} = n \ket{n}$ ($n = 0, \pm 1, \pm 2, \ldots$), and the operator $\hat{U}^{+} = \exp( + i \hat{\theta})$ [$\hat{U}^{-} = \exp( - i \hat{\theta})$] raises (lowers) the charge of a state by one, $\hat{U}^{\pm} \ket{n} = \ket{n \pm 1}$. With a truncation $|n|_{\rm{max}} = S$, $\hat{U}^{\pm}$ and $\hat{S}^z$ have the following commutation relations
\begin{eqnarray}
\left[\hat{U}^{+}, \hat{U}^{-}\right] &=& \hat{D}, \label{eq:upumcommute} \\ \left[\hat{S}^z, \hat{U}^{\pm}\right] &=& \pm \hat{U}^{\pm} \label{eq:szupmcommute},
\end{eqnarray}
where $\bra{n'}\hat{D}\ket{n} = \delta_{n',n}\delta_{n,2S+1}-\delta_{n',n}\delta_{n,-2S-1}$, which means the matrix elements of $\hat{D}$ are all zero except the most upper-left one ($\bra{2S+1}\hat{D}\ket{2S+1} = 1$) and the most lower-right one ($\bra{-2S-1}\hat{D}\ket{-2S-1}=-1$).

The Hamiltonian in Eq. \eqref{eq:any-spin-ham-charge} has an explicit global U(1) symmetry, so the total charge (or magnetization) is a conserved quantum number for any spin truncation, as it should be for the O(2) model as the matter fields of compact sQED. The phase diagrams in Refs. \cite{PhysRevB.67.104401, PhysRevB.87.235106} show that Eq. \eqref{eq:any-spin-ham-charge} with $Y=0, X \neq 0$ is gapless for $S = 1, 2$. It is expected that the $X$ term is gapless for any $S$ in the thermodynamic limit, which may drive a quantum phase transition from a gapped phase to a gapless phase belonging to the BKT type. In the limit $Y \gg X$, Eq. \eqref{eq:any-spin-ham-charge} is equivalent to a simple Bose Hubbard model that can be prepared in the cold atom experiment to study the scaling of entanglement entropy in the superfluid phase with incommensurate charge filling \cite{PhysRevD.96.034514, PhysRevA.96.023603}.

For $S=1$, early works indicate that the quantum phase transition in Eq. \eqref{eq:any-spin-ham-charge} belongs to the BKT type \cite{PhysRevB.16.1153, PhysRevB.23.4698, Jullien_1981, DENNIJS1982273, PhysRevB.30.1629, doi:10.1063/1.333675, PIRES2007387}. However, some of these works also provide clues that there exist some differences from BKT. Luther and Scalapino \cite{PhysRevB.16.1153} asserted that the correlation-function exponent $\eta = 1/\sqrt{8}$ is inconsistent with BKT. Their approach was revisited by den Nijs \cite{DENNIJS1982273} who obtained $\eta = 1/4$, consistent with BKT. Reference \cite{Jullien_1981} concluded that there is an essential singularity at the transition point but did not extract a reliable $\sigma$ assuming $\xi \sim \exp\left[b/(Y-Y_c)^{\sigma}\right]$. Reference \cite{PhysRevB.23.4698} obtained $\sigma = 0.9(3)$ and Ref.~\cite{doi:10.1063/1.333675} obtained accurate values for the dynamic exponent $z=1.00(1)$ and for the $\eta = 0.26(2)$ that are consistent with BKT, but was not successful in extracting reliable $\sigma$, either. With the development of LS techniques to locate quantum phase transitions \cite{Nomura_1995, Nomura_1998}, the ground-state phase diagram of the spin-$1$ $XXZ$ chain with single-ion anisotropy $D$ was mapped out in Ref.~\cite{PhysRevB.67.104401}. Our Hamiltonian \eqref{eq:any-spin-ham-charge} corresponds to the Hamiltonian of Ref.~\cite{PhysRevB.67.104401} with $J_z = 0$, where $J_z$ is the coupling of $S^z_l S^z_{l+1}$. It resides at the boundary between the gapless $XY$ phase and the gapped odd Haldane phase for small $Y$ and is in a gapped phase (large-$D$ phase in Ref.~\cite{PhysRevB.67.104401}) for large $Y$. The phase-transition point $Y_c$ is the intersection of three critical lines: the BKT line separating the $XY$ phase and the large-$D$ phase, the BKT line separating the $XY$ phase and the odd Haldane phase, and the Gaussian line separating the odd Haldane phase and the large-$D$ phase. So the quantum phase transition in Hamiltonian \eqref{eq:any-spin-ham-charge} with $S = 1$ should be an infinite-order Gaussian transition from a gapped phase to a gapless BKT critical line. This kind of transition is on one of the $y_0=\pm y_\phi$ lines of the RG equations for the sine-Gordon (SG) model (see Appendix~\ref{apdx:sinegordan}), where there is an inherent SU(2) symmetry \cite{PhysRevD.12.1684}. On the lines $y_0=\pm y_\phi$, the correlation length diverges as $Y \rightarrow Y_c$ with an essential singularity of the form $\xi \sim (Y-Y_c)^{-1/2}\exp\left[b/(Y-Y_c)\right]$ \cite{PhysRevLett.79.3214, PhysRevB.60.7850, PhysRevB.61.16377}, the same as for the spin-gap phase transition in the Hubbard model \cite{ovchinnikov1970excitation, 10.1143/PTP.48.2171, PhysRevB.60.7850}, instead of as $\xi \sim \exp(b/\sqrt{Y-Y_c})$ \cite{Kosterlitz_1974} as is the case for the BKT transition. The connection of the $S = 1$ case to the Hubbard model can be seen by writing the spin-$1$ operators as an addition of two spin-$1/2$ operators, $\hat{S}^\alpha = \hat{r}^\alpha + \hat{t}^\alpha$, such that the Hamiltonian \eqref{eq:any-spin-ham-charge} with $S = 1$ can be written in the form \cite{DENNIJS1982273}
\begin{eqnarray}
\label{eq:s1split2shalves}
\nonumber && \hat{H}_{c}(S = 1) = Y \sum_{l=1}^{L} \hat{r}_{l}^z \hat{t}_{l}^z \\ \nonumber  &&- \frac{X}{4} \sum_{l = 1}^{L-1} (\hat{r}_{l}^+ \hat{r}_{l+1}^- + \hat{r}_{l}^- \hat{r}_{l+1}^+ + \hat{t}_{l}^+ \hat{t}_{l+1}^- + \hat{t}_{l}^- \hat{t}_{l+1}^+) \\ &&- \frac{\lambda}{4} \sum_{l = 1}^{L-1} (\hat{r}_{l}^+ \hat{t}_{l+1}^- + \hat{t}_{l}^- \hat{r}_{l+1}^+ + \hat{t}_{l}^+ \hat{r}_{l+1}^- + \hat{r}_{l}^- \hat{t}_{l+1}^+)
\end{eqnarray}
with $\lambda = X$. Using the Jordan-Wigner transformation, the first two terms of Eq. \eqref{eq:s1split2shalves} can be exactly mapped to the Hubbard model, which exhibits SU(2) symmetry. The last term is the interspecies hopping term and breaks the Hubbard SU(2) symmetry. It was pointed out by den Nijs \cite{DENNIJS1982273} that there should be a line of BKT transition points emerging from the point at $\lambda = 0$ to the point at $\lambda = X$. The fact that the system is on a BKT line for any $Y < Y_c$ indicates that the model should have a hidden SU(2) symmetry. This hidden SU(2) symmetry emerges as the spin-$1/2$ Heisenberg chain for large negative $Y$ \cite{refId0}, where $S^z = 0$ states are gapped out and $\ket{\pm 1}$ states act as spin-up and spin-down states in the spin-$1/2$ Heisenberg chain. The new effective spin-$1/2$ operators are $\tilde{S}^z = (1/2)S^z, \tilde{S}^+ = (1/2)S^+ S^+, \tilde{S}^- = (1/2)S^-S^-$. As the ground state is not likely to have another symmetry breaking in the gapless phase, we expect that the SU(2) symmetry of the Heisenberg chain at large negative $Y$ smoothly connects to the hidden SU(2) symmetry up to the phase-transition point $Y = Y_c > 0$.

The hidden SU(2) symmetry in $\hat{H}_c(S=1)$ ensures that at the transition point $Y_c$ corresponding to $y_0 = y_\phi = 0$ in the SG model, the coupling constant of the marginal operators is zero and the multiplicative logarithmic corrections to the correlation function vanishes, but the critical exponents are expected to be the same as BKT. Reference \cite{PhysRevB.67.104401} does not give the numeric value of this phase-transition point, but it is around $0.35$ from the phase diagram. In Ref.~\cite{Jullien_1981}, $Y_c = 0.4$, Ref.~\cite{doi:10.1063/1.333675} gives $Y_c = 0.50(5)$, Ref.~\cite{PIRES2007387} gives $Y_c = 0.475$, and Ref.~\cite{Langari_2013} gives $Y_c = 0.347$. In this work, we give a much more accurate number by performing LS with $L$ up to $21$. Note that, because the rotation of the spin around $z$ axis by $\pi$ ($S^x \rightarrow -S^x, S^y \rightarrow -S^y, S^z \rightarrow S^z$) on odd or even sites is equivalent to a change of sign of $X$ ($X \rightarrow -X$), the phase diagram does not depend on the sign of $X$. For $S \ge 2$, the $XXZ$ chain with single-ion anisotropy has a very different phase diagram \cite{PhysRevB.87.235106}. A key difference is that the Haldane phase is pushed to positive $J_z$ values. So our Hamiltonian \eqref{eq:any-spin-ham-charge} with $S \ge 2$ truncation is deeply inside the gapless $XY$ phase for small $Y$. Decreasing $Y$ drives the system to go across the BKT critical line, so the phase transition is of a true BKT type.

%%%%%%%%%%%%%%%%%%%%%%%%%%%%%%%%%%%%%%%%%%%%%%%%%%%%%
\subsection{$\hat{U}^{\pm}$ and $\hat{S}^{\pm}$} \label{subsec:upmandspm}
%%%%%%%%%%%%%%%%%%%%%%%%%%%%%%%%%%%%%%%%%%%%%%%%%%%%%
Another kind of truncation is to replace $\hat{U}^{\pm}$ operators by the spin ladder operators $\hat{S}^{\pm} / \sqrt{S(S+1)}$, which are used in the quantum link models of LGTs \cite{CHANDRASEKHARAN1997455, PhysRevD.60.094502, PhysRevLett.124.123601}. $\hat{S}^{\pm}$ and $\hat{S}^z$ satisfy $[\hat{S}^+, \hat{S}^-] = 2 \hat{S}^z$ different from Eq. \eqref{eq:upumcommute} for $S \ge 2$ and $[\hat{S}^z, \hat{S}^{\pm}] = \pm \hat{S}^{\pm}$ which is the same as Eq. \eqref{eq:szupmcommute} for any $S$. $\hat{U}^{\pm}$ and $\hat{S}^{\pm}$ operators can be related by the following equation,
\begin{eqnarray}
\label{eq:oprelation}
\hat{U}^{\pm} = u_0 \hat{S}^{\pm} + \sum_{q=1}^{S-1} u_q \left(\hat{S}^z\right)^q \hat{S}^{\pm} \left(\hat{S}^z\right)^q.
\end{eqnarray}
The coefficients $u_q$ can be found by solving the linear equations
\begin{eqnarray}
\label{eq:auequalb}
\boldsymbol{A} \boldsymbol{u} = \boldsymbol{b}, 
\end{eqnarray}
where $\boldsymbol{A}$ is an $S \times S$ matrix with elements $A_{ij} = [(S-i)(S-i-1)]^j$ ($i, j = 0, 1, \ldots, S-1$), $\boldsymbol{u} = (u_0, u_1, ..., u_{S-1})^{\boldsymbol{T}}$, and $\boldsymbol{b} = (b_0, b_1, \ldots, b_{S-1})^{\boldsymbol{T}}$ with $b_j = 1/\sqrt{S(S+1) - (S-j)(S-j-1)}$. For the first few spin truncations,
\begin{eqnarray}
\label{eq:us1ladderexample}
\hat{U}^{\pm} &=& \frac{1}{\sqrt{2}} \hat{S}^{\pm}, S = 1; \\ 
\label{eq:us2ladderexample}
\hat{U}^{\pm} &=& \frac{1}{\sqrt{6}} \hat{S}^{\pm} + \left(\frac{1}{4} - \frac{\sqrt{6}}{12} \right) \hat{S}^z \hat{S}^{\pm} \hat{S}^z, S = 2; \\
\label{eq:us3ladderexample}
\nonumber \hat{U}^{\pm} &=& \frac{1}{\sqrt{12}} \hat{S}^{\pm} + \left(-\frac{\sqrt{6}}{72} - \frac{\sqrt{3}}{9} + \frac{3\sqrt{10}}{40} \right) \hat{S}^z \hat{S}^{\pm} \hat{S}^z \\ &+& \left(\frac{\sqrt{6}}{144} + \frac{\sqrt{3}}{72} - \frac{\sqrt{10}}{80} \right)(\hat{S}^z)^2 \hat{S}^{\pm} (\hat{S}^z)^2, S = 3.
\end{eqnarray}
In particular, $u_0 = 1/\sqrt{S(S+1)}$, which normalizes the amplitude of raising (lowering) $\ket{0}$ to $\ket{+1}$ ($\ket{-1}$). The matrix elements of $u_0 \hat{S}^{\pm}$ are $\delta_{i, j \pm 1} \sqrt{1-j(j \pm 1)/S(S+1)}$. For infinite $S$, the nonzero matrix elements of $u_0 \hat{S}^{\pm}$ at finite $i, j$ are all equal to $1$. In other words, $\hat{U}^{\pm} = \hat{S}^{\pm}/\sqrt{S(S+1)}$ for $S = 1$ or $S \rightarrow \infty$. For finite $S \ge 2$, the difference between $\hat{U}^{\pm}$ and $u_0 \hat{S}^{\pm}$ is small: we expect the two kinds of truncation schemes to have the same type of quantum phase transitions. The fine structure of the linear system in Eq. \eqref{eq:auequalb} is discussed in Appendix~\ref{apdx:linearequations}, where we show that the magnitude of $u_q$ decays exponentially with the index $q$.

%%%%%%%%%%%%%%%%%%%%%%%%%%%%%%%%%%%%%%%%%%%%%%%%%%%%
\subsection{Level spectroscopy}\label{subsec:modells}
%%%%%%%%%%%%%%%%%%%%%%%%%%%%%%%%%%%%%%%%%%%%%%%%%%%%
The LS method for the BKT transition is based on detailed analysis of energy excitations using conformal field theory (CFT). For a pure Gaussian model with PBC, each excitation classified by quantum numbers $\mathbf{j} = (M, k, P)$, where $M$ is the total charge or magnetization, $k$ is the wavenumber, and $P$ is the parity, has the energy gap $\Delta E_{\mathbf{j}}$ and the scaling dimension $x_{\mathbf{j}}$ that are related by $\Delta E_{\mathbf{j}} = 2\pi \nu x_{\mathbf{j}} / L$, where $\nu$ is the spin-wave velocity. In the neighborhood of a BKT critical line, the scaling dimensions of the marginal operators deviate from $2$ in different ways, which may cause a level crossing. As shown in Refs. \cite{Nomura_1995, Nomura_1998} for the BKT transition without symmetry breaking, one of the proper choices is the level crossing between excitations $(M=\pm 4, k = 0, P = 1)$ and $(M = 0, k = 0, P = 1)$. In the effective SG theory with coupling constants $y_0, y_\phi$ (see Appendix \ref{apdx:sinegordan}), the renormalized scaling dimensions for these two excitations are $2-y_0(l)$ and $2-y_0(l)(1+4t/3)$, respectively, where $t$ is the distance to the BKT critical line. For $S = 1$, decreasing $Y$ drives the system into a BKT line at $t = 0$, and we expect to see an exact degeneracy in these energy levels for any $Y$ small enough and any finite $L$. The phase-transition point corresponds to the multicritical point $y_0 = y_\phi = 0$ in the SG theory. Another method for spin-$1$ truncation is to apply twisted boundary conditions (TBCs), $\hat{S}^z_{L+1} = \hat{S}^z_1, \hat{U}^{\pm}_{L+1} = - \hat{U}^{\pm}_1$ \cite{PhysRevLett.76.4038, doi:10.1143/JPSJ.66.3944, doi:10.1143/JPSJ.66.3379}, and study the level crossing between the ground-state energy in sector $(M=0, P=1)$ and the ground-state energy in sector $(M=0, P=-1)$. This method is mainly used in Ref. \cite{PhysRevB.67.104401} to locate the Gaussian line between the odd Haldane phase and the large-$D$ phase. Since at small $Y$, the Hamiltonian \eqref{eq:any-spin-ham-charge} with $S = 1$ truncation is on the phase boundary of the Haldane phase, TBCs still induce a Haldane gap for finite-size systems. The same mechanism allows one to consider an odd number of sites with PBCs, and study the level crossing between the ground-state energy in the sector $(M=0, k=0, P=1)$ and that in the sector $(M=0, k=0, P=-1)$. This method is equivalent to the method in Ref. \cite{Langari_2013} where the discontinuity of the ground-state expectation value of the permutation operator $P_{i, j} = \vec{S}_i \cdot \vec{S}_j + (\vec{S}_i \cdot \vec{S}_j)^2 - 1$ is used to locate the same Gaussian line. For $S \ge 2$, the phase transition is of a true BKT type and there is true level crossing between the excitations $(M=\pm 4, k = 0, P = 1)$ and $(M = 0, k = 0, P = 1)$.

%%%%%%%%%%%%%%%%%%%%%%%%%%%%%%%%%%%%%%%%%%%%%%%%%%%%
\subsection{Gap scaling} \label{subsec:modelgs}
%%%%%%%%%%%%%%%%%%%%%%%%%%%%%%%%%%%%%%%%%%%%%%%%%%%%
As any phase transition happens at the place of closing the energy gap, the energy gap between the lowest two levels is a natural universal tool for quantum phase transitions. Unlike LS, this method does not require a prior CFT analysis of the target model and the inverse of the energy gap describes the divergent behavior of the correlation length. To apply this method to our model, the first idea is to extrapolate the energy gap to the thermodynamic limit, and fit the data with $A \left(Y-Y_c\right)^{1/2}\exp{\left[-b/(Y-Y_c)\right]}$ for $S = 1$ and $A \exp{(-b/\sqrt{Y-Y_c})}$ for $S \ge 2$ \cite{PhysRevLett.76.2937}. This is usually difficult and requires precise manipulations of the extrapolation procedure. A more stable way is to use the following ansatz for the scaling of the energy gap in the vicinity of the phase transition \cite{PhysRevB.84.115135, PhysRevA.87.043606, PhysRevB.91.165136}:
\begin{eqnarray}
\label{eq:gapscaling}
L \Delta E \left[ 1 + g\left( L \right) \right] = F\left( \frac{\xi}{L} \right),
\end{eqnarray}
where the correlation length
\begin{eqnarray}
\xi \sim (\Delta E)^{-1} \sim \begin{cases}
     \left(Y-Y_c\right)^{-1/2} e^{b/(Y-Y_c)}, & S = 1\\
    e^{b/\sqrt{Y-Y_c}}, & S \ge 2
\end{cases}
\end{eqnarray}
near the phase transition in the gapped phase. $F(\xi / L)$ is a universal scaling function and $g(L)$ is a correction term depending on the size of the system. The leading behavior of $g(L)$ is $ 1/\left[2\ln(L) + C\right]$ from Weber and Minnhagen \cite{PhysRevB.51.6163}. We can also include higher-order corrections and take $g(L) = 1/\left[ 2\ln(L) + C + \ln(C/2+\ln(L))\right] + A / \ln^2(L)$ \cite{Hasenbusch_2005(1), Hasenbusch_2005(2), Hasenbusch_2008, PhysRevE.87.032105, Hsieh_2013} to further decrease the error. The goal of this method is to find the best data collapse of the rescaled energy gap $\Delta E_s = L \Delta E \left[ 1 + g\left( L \right) \right]$ near the phase-transition point in the parameter space $(Y_c, b, C, A)$. The universal function $F(\xi / L)$ is approximated by an arbitrarily high degree polynomials of the variable $x_L = \ln(L/\xi)$. We show that the best data collapse is found at $C \rightarrow \infty$, which implies that logarithmic corrections are highly suppressed. The phase-transition points obtained from this method without $g(L)$ differ from those from LS only at the third decimal place.

%%%%%%%%%%%%%%%%%%%%%%%%%%%%%%%%%%%%%%%%%%%%%%%%%%%%
\subsection{Parameters in numerical algorithms} \label{subsec:modeltrgdmrg}
%%%%%%%%%%%%%%%%%%%%%%%%%%%%%%%%%%%%%%%%%%%%%%%%%%%%
 
The tensor contraction in the path integral Eq. \eqref{eq:o2pathintegralbessel} can be calculated efficiently by the higher-order tensor renormalization-group (HOTRG) method \cite{PhysRevB.86.045139}. The local observables such as the magnetization $M = \langle \cos(\theta) \rangle$ can be calculated using the impure tensor method \cite{MORITA201965, PhysRevD.100.054510}. When contracting the four-rank tensor in Eq. \eqref{eq:o2pathintegralbessel}, the tensorial bond dimension grows exponentially. We restrict the maximal bond dimension to be $D_{\rm{bond}}$ in the calculation. The maximal lattice size we use is $V = 2^{24} \times 2^{24}$, where the calculated quantities converge within $12$ significant numbers such that we are effectively in the thermodynamic limit. The maximal tensorial bond dimension is set to be $D_{\rm{bond}} = 60$ for $S = 1$ and $D_{\rm{bond}} = 42$ for $S \ge 2$ to ensure that the dependence of the results on the bond dimension is small. 

\begin{figure}
  \centering
    \includegraphics[width=0.48\textwidth]{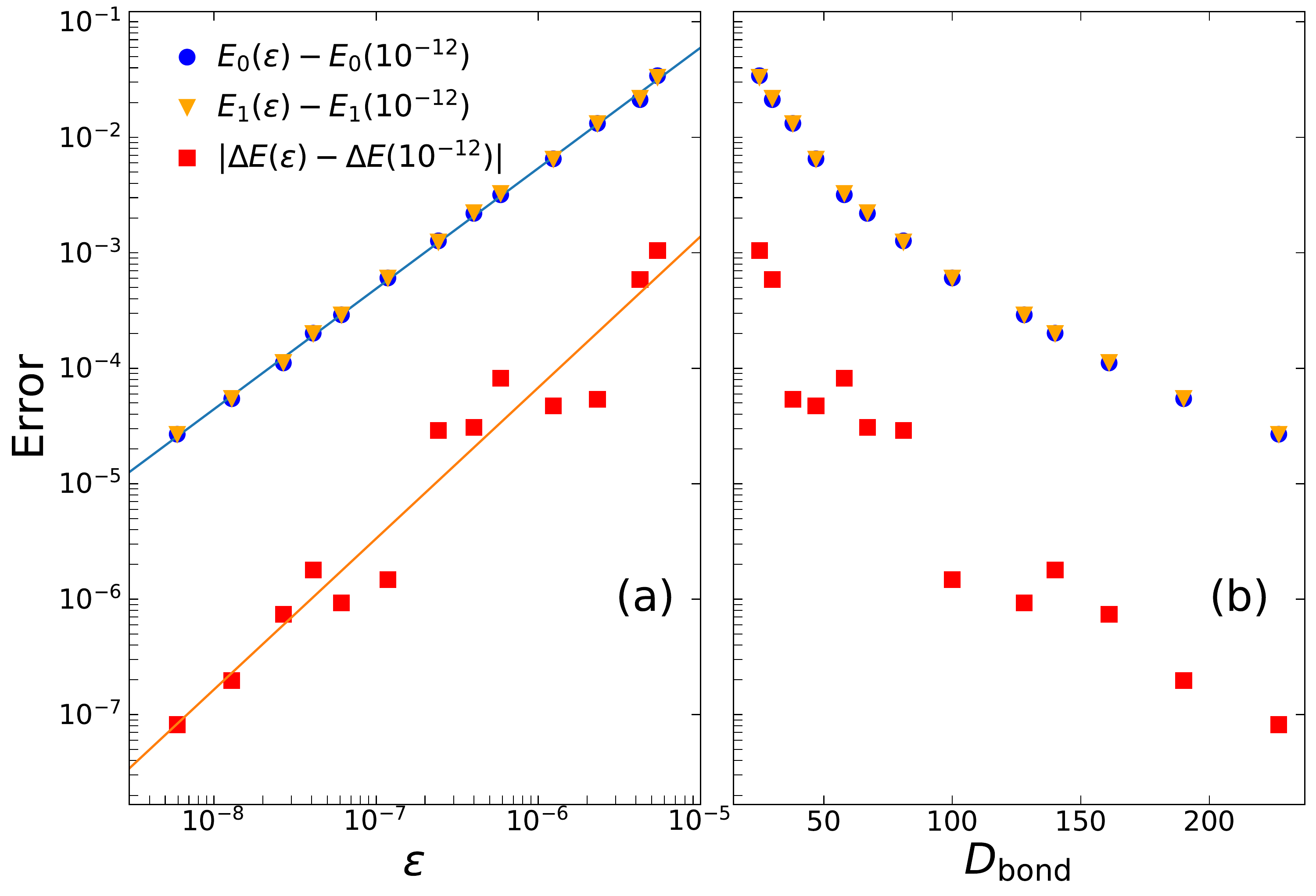}
    \caption{\label{fig:dmrgparamscaling}The dependence of the error in the lowest two eigenenergies and the energy gap on (a) the truncation error and (b) the bond dimension in DMRG calculations. The error in energy is obtained by subtracting the results for $\epsilon = 10^{-12}$ from those for larger truncation errors. The results are for $\hat{S}^{\pm}$ operators with $S = 2, L = 512, Y = 0.94$. Linear fits give $E_0(\epsilon) - E_0 = 10^{3.99(6)}\epsilon^{1.043(9)}$ and $|\Delta E (\epsilon) - \Delta E| = 10^{3.7(7)}\epsilon^{1.31(10)}$.
    }
  \end{figure}

For the Hamiltonian approach, we use the finite-size DMRG \cite{PhysRevLett.69.2863, PhysRevB.48.10345, SCHOLLWOCK201196} algorithm with matrix product state (MPS) \cite{PhysRevLett.75.3537} optimization to calculate the energy gap between the lowest two levels. The calculations are performed with the ITENSOR C++ Library \cite{itensor}. We increase the number of Schmidt states gradually during the finite-size sweeping procedure until the truncation error $\epsilon$ is less than $10^{-10}$, which requires the largest bond dimension for the data used in this paper to be around $665, 782, 698, 601, 537$ for $S = 1,2,3,4,5$ respectively. The number of sweeps is large enough for the difference in the energy between the last two sweeps to be less than $10^{-12}$. The smallest energy gap we calculate is of order $10^{-3}$ with a typical error $\approx 10^{-8}$ estimated by comparing the results to those for $\epsilon = 10^{-12}$. The largest bond dimension for $\epsilon = 10^{-12}$ is around $1400$. By subtracting the results for $\epsilon = 10^{-12}$ from those for larger $\epsilon$, we show the dependence of the error in eigenenergies and the energy gap on the truncation error and the bond dimension in Fig.~\ref{fig:dmrgparamscaling}. One can see that, in the logarithmic scale, the error is linear with $\epsilon$, which means that the error is a power-law scaling function of $\epsilon$. The power for the energy gap is $1.31(10)$, larger than $1.043(9)$ for the eigenenergies. The errors in the lowest two eigenenergies are almost the same, thus the energy gap has a significantly smaller error. It is also seen that the error decreases exponentially with bond dimension, which is consistent with the results in Ref.~\cite{PhysRevB.48.10345}. These observations guarantee that our results are accurate enough so that the error from DMRG is negligible in the following analysis. We set $X = 1$ in all the calculations for the Hamiltonian unless otherwise specified.

%%%%%%%%%%%%%%%%%%%%%%%%%%%%%%%%%%%%%%%%%%%%%%%%%%%%%%
\section{Results}\label{sec:results}
%%%%%%%%%%%%%%%%%%%%%%%%%%%%%%%%%%%%%%%%%%%%%%%%%%%%%%

%%%%%%%%%%%%%%%%%%%%%%%%%%%%%%%%%%%%%%%%%%%%%%%%%%%%%
\subsection{Lagrangian: magnetic susceptibility} \label{subsec:lagmagneticsus}
%%%%%%%%%%%%%%%%%%%%%%%%%%%%%%%%%%%%%%%%%%%%%%%%%%%%%
\begin{figure}
  \centering
    \includegraphics[width=0.48\textwidth]{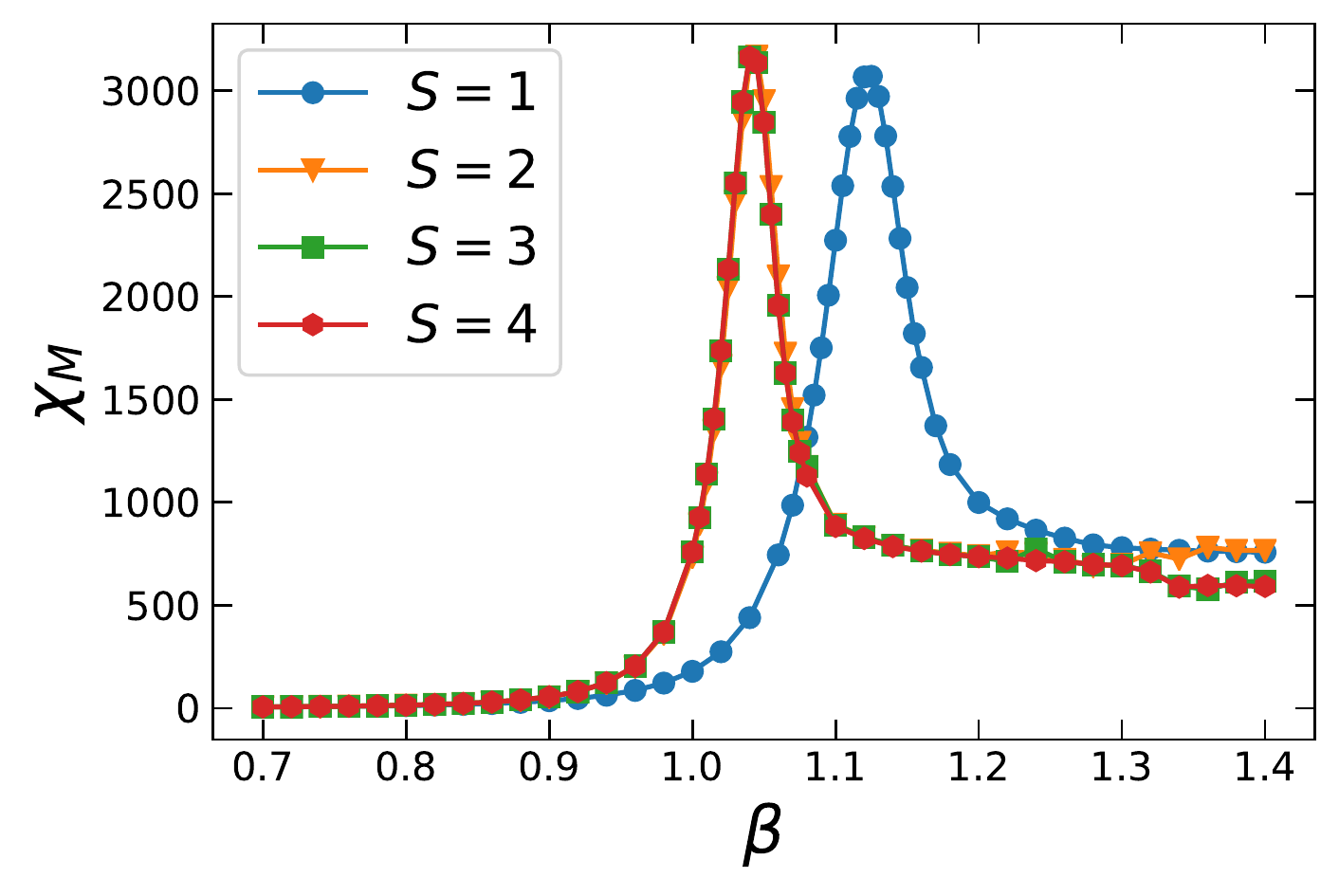}
    \caption{\label{fig:msus1-4inf}The magnetic susceptibility of O(2) model with $h = 4 \times 10^{-5}$ as a function of inverse temperature $\beta$ for different spin truncations. The data for $S = 3,4$ are on top of each other. The volume of the lattice is $V = 2^{24} \times 2^{24}$. The tensorial bond dimension is $D_{\rm{bond}} = 40$.
    }
  \end{figure} 
  
Since the charge representation seems to preserve the infinite-order quantum phase transition from gapped to gapless phase for any spin truncation (Gaussian for $S = 1$ and BKT for $S \ge 2$), a natural question is how the transition point $\beta_c$ or $Y_c$ changes with spin truncation. The first step is to check the magnetic susceptibility $\chi_M = (1/V)\partial^2 \ln Z / \partial h^2$ in the path integral formulation in Euclidean space-time. In practice, the magnetic susceptibility at $h$ is calculated by symmetric numerical differentiation
\begin{eqnarray}
\chi_M\left(h\right) = \frac{M\left(h+\delta h\right) - M\left(h-\delta h\right)}{2\delta h},
\end{eqnarray}
where the magnetization $M(h)$ is calculated by HOTRG with impure tensor method \cite{PhysRevD.100.054510, MORITA201965}. The magnetic susceptibility $\chi_M$ as a function of $\beta$ is presented in Fig.~\ref{fig:msus1-4inf}. At weak external field $h = 4\times 10^{-5}$, the peak of $\chi_M$ for spin-$1$ truncation is around $\beta = 1.12$, while for spin-$2$ and above, the peaks are all around $\beta = 1.04$. The values of $\chi_M$ at spin-$2$ truncation already effectively converges to their large-$S$ value. At small $\beta$, $\chi_M$ is close to zero because it is in the disordered gapped phase at high temperature, while it has a high plateau at large $\beta$ across the peak. Both the peak and the height of the plateau diverges when the external field approaches zero. These facts indicate that for all spin truncations, the low temperature phase is a gapless phase with infinite correlation length. The results agree with the picture of the BKT transition in the classical O(2) model.
\begin{figure}
  \centering
    \includegraphics[width=0.48\textwidth]{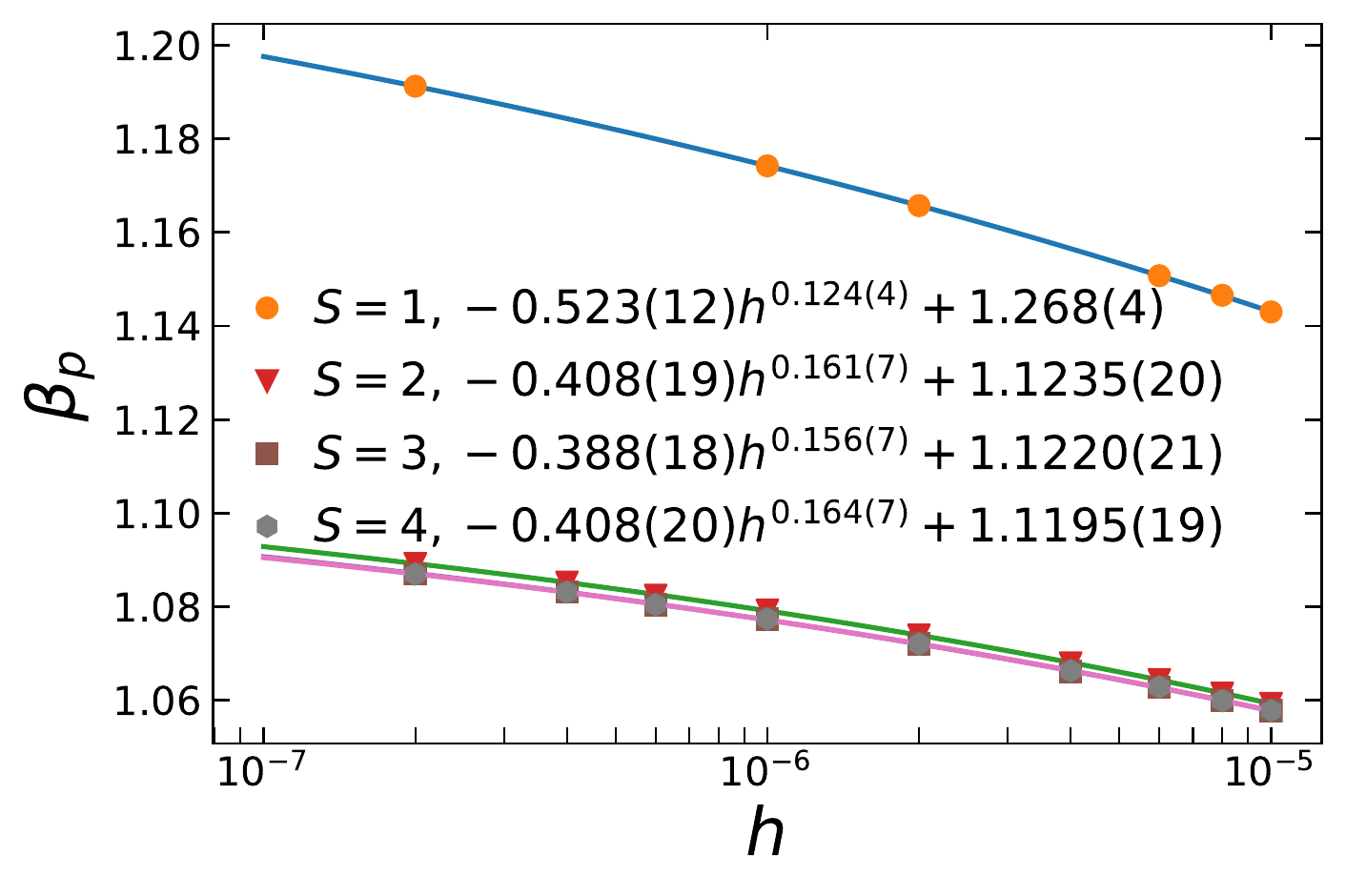}
    \caption{\label{fig:msus1-4extrap}Power-law extrapolations of the peak positions of $\chi_M$ to zero external field for truncations $S = 1, 2, 3, 4$. The tensorial bond dimension $D_{\rm{bond}} = 60$ for $S = 1$, and $D_{\rm{bond}} = 42$ for $S \ge 2$. The data for $S = 3$ and $S = 4$ have invisible difference and stay on top of each other. The extrapolated results are consistent with the Monte Carlo result $\beta_c = 1.11996(6)$ for the O(2) model in Ref.~\cite{doi:10.1143/JPSJ.81.113001}.
    }
  \end{figure} 

Reference \cite{PhysRevE.89.013308} uses HOTRG with $D_{\rm{bond}} = 40$ to calculate the magnetic susceptibility and obtains $T_c = 0.8921(19)$ for the O(2) model, consistent with other works. For spin-$1$ truncation, the result is more sensitive to the bond dimension in HOTRG. We test the bond dimension and find that using $D_{\rm{bond}} = 60$ for $S = 1$ and $D_{\rm{bond}} = 42$ for $S = 2, 3, 4$ is good enough for $\chi_M$ to converge within $0.1\%$ error. Performing the same procedure as described in Ref.~\cite{PhysRevE.89.013308}, we extrapolate the position of the peak of $\chi_M$ to zero external field by a power law. Figure \ref{fig:msus1-4extrap} shows that the extrapolated critical inverse temperatures $\beta_c = 1.268(4), 1.1235(20), 1.1220(21), 1.1195(19)$ for $S = 1, 2, 3, 4$ respectively. Within uncertainties, $\beta_c$ already converges at $S = 2$. It is expected that the phase-transition point converges at least exponentially fast with spin truncation, which is also confirmed in the following sections using the Hamiltonian approach. For the charge representation with $S \ge 2$, where there should be a BKT transition, the exponent in the power law $\beta_p - \beta_c = a h^b$ is close to the value $b = 0.1768$ obtained in Ref.~\cite{PhysRevE.89.013308} and $b = 0.162(1)$ obtained in Ref.~\cite{Jha_2020}, while it is very different for $S = 1$, indicating a different type of phase transition.

\begin{figure}
  \centering
    \includegraphics[width=0.48\textwidth]{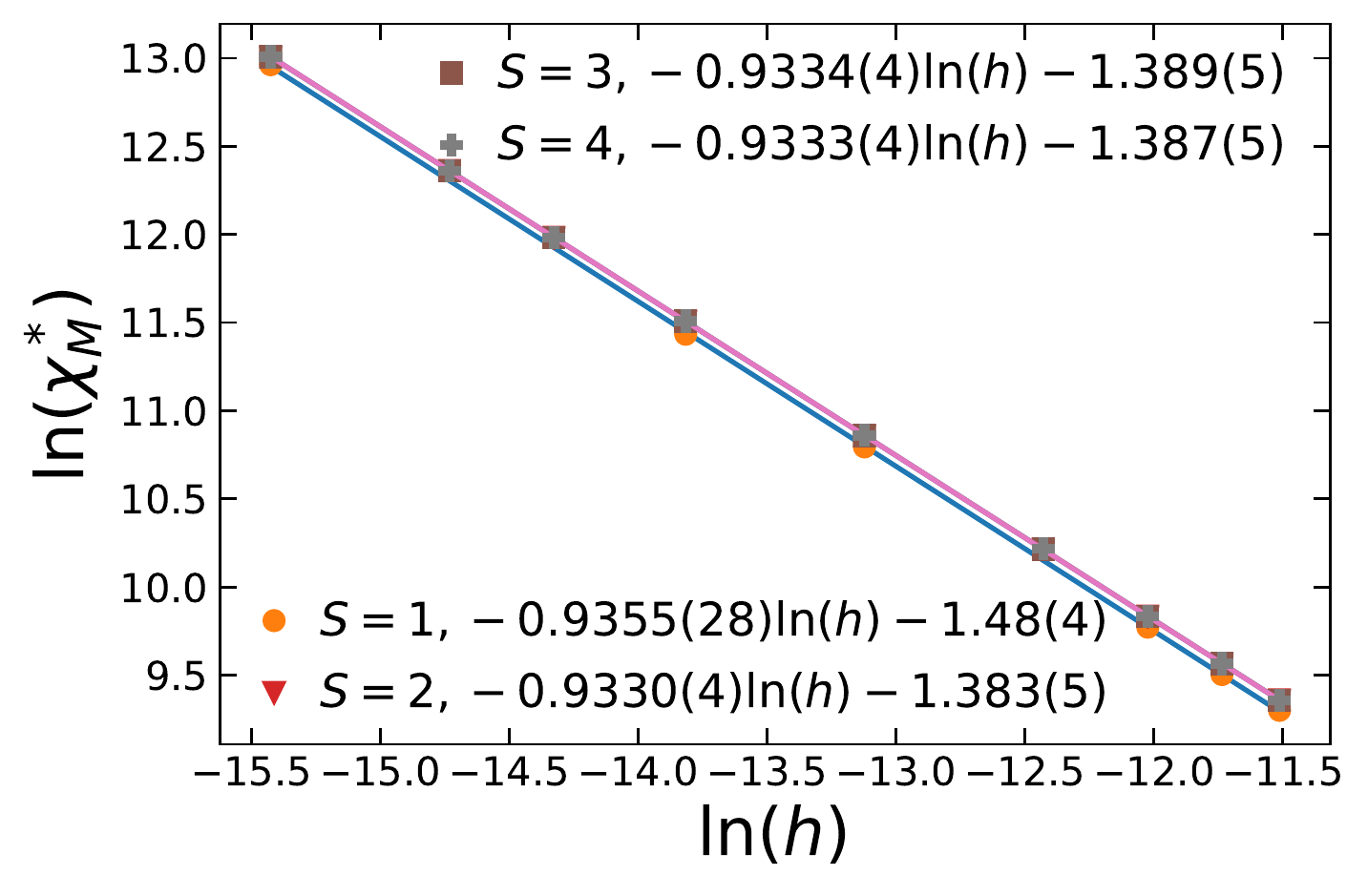}
    \caption{\label{fig:msusdeltaexponent}The maximal magnetic susceptibility as a function of external field for $S = 1, 2, 3, 4$. The data for $S = 2, 3, 4$ have invisible difference and stay on top of each other. The $\delta$ exponents are found to be $15.5(7), 14.93(9), 15.02(9), 14.99(9)$ for $S = 1, 2, 3, 4$ respectively. 
    }
  \end{figure}

Another question is whether the smallest spin truncation changes the magnetic critical exponent $\delta$ in the power law of magnetization $M \sim h^{1/\delta}$. Instead of doing a curve fit for $M(h)$, we perform a linear fit for the plot of $\ln(\chi^*_{M})$ versus $\ln(h)$, where $\chi^*_{M}$ is the peak height of $\chi_M$. The slope of the linear fit is expected to be $1/\delta - 1 = -14/15$ for the BKT transition. Figure \ref{fig:msusdeltaexponent} depicts this procedure. It is found that the fitted slopes are $-0.9355(28), -0.9330(4), -0.9334(4), -0.9333(4)$ for $S = 1,2,3,4$ respectively, which gives the magnetic critical exponent $\delta = 15.5(7), 14.93(9), 15.02(9), 14.99(9)$. All the values are consistent with the predicted value for the BKT transition $\delta = 15$. As discussed before, the phase transition for the charge representation with $S = 1$ should be an infinite-order Gaussian transition at the end of a BKT line, the agreement on the $\delta$ exponent between $S = 1$ and $S \ge 2$ is consistent with this picture.
%%%%%%%%%%%%%%%%%%%%%%%%%%%%%%%%%%%%%%%%%%%%%%%%%%%%%%
\subsection{Hamiltonian: Level spectroscopy} \label{subsec:hamls}
%%%%%%%%%%%%%%%%%%%%%%%%%%%%%%%%%%%%%%%%%%%%%%%%%%%%%%
In Sec.~\ref{subsec:modells}, we mention three ways to perform LS to locate the phase-transition point for $S = 1$. We first discuss the TBC method. As shown in Refs.~\cite{Nomura_1995, PhysRevLett.76.4038, doi:10.1143/JPSJ.66.3944, PhysRevB.67.104401}, for small $Y$ and $X = -1$, the ground state is on the boundary of the odd Haldane phase with $P = -1, T = -1$, where $T$ is the spin reversal symmetry. For large $Y$ and $X = -1$, the ground state is in the large-$D$ phase, where $P = 1, T = 1$. There must be a level crossing between the two parity sectors. Note that for $X=-1$, the level crossing only exists for even total number of sites. This is because if $X = -1$ and TBC is applied, the bulk spins are coupled with positive coefficients and the edge spins are coupled with a negative coefficient. On the boundary of the odd Haldane phase at finite system size, the bulk spins form valence bonds which are singlets with $P = -1, T = -1$, while the edge spins form a triplet with $P = 1, T = 1$. The number of sites needs to be even to form odd number of singlets such that $P = -1, T = -1$ for the whole system. For odd number of sites and $X = -1$, there is no level crossing between the two parity sectors. However, when using $X = 1$ with TBC, an odd number of total sites can form a singlet with $P = -1, T = -1$ for the edge spins, while the bulk spins form triplets with $P = 1, T = 1$ for small $Y$. Therefore level crossings exist for all even and odd number of sites.

\begin{figure}
  \centering
    \includegraphics[width=0.48\textwidth]{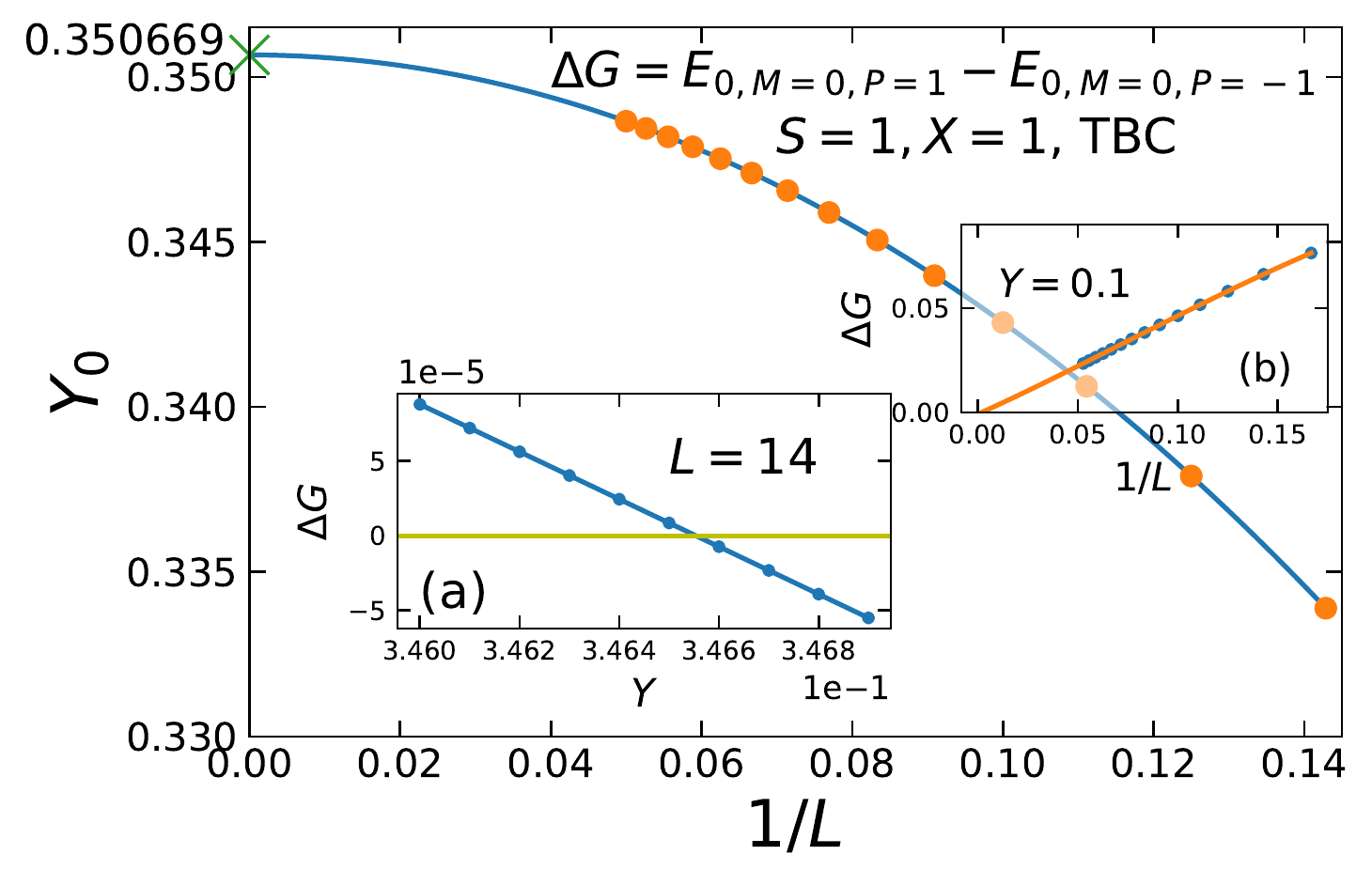}
    \caption{\label{fig:x1tbclevel}The extrapolation procedure of finite size $Y_c$ for $S = 1, X = 1$, TBCs. The finite size $Y_c$ for $L$ up to $20$ is found by locating the position of the level crossing between the ground-state energy of the sector $M=0, P=1$ and the ground-state energy of the sector $M=0, P=-1$. The extrapolated $Y_c = 0.3506694(3)$. The inset (a) shows the level crossing near $Y_0$ for $L = 14$. The inset (b) shows the energy gap as function of $1/L$ for $Y = 0.1$.
    }
  \end{figure}

In practice, we calculate the ground state in the sector $M = 0, P = -1$ with energy $E_{0, M = 0, P = -1}$ and the ground state in sector $M = 0, P = 1$ with energy $E_{0, M = 0, P = 1}$, and locate $Y_0$ where the energy gap $\Delta G = E_{0, M = 0, P = 1} - E_{0, M = 0, P = -1}$ changes sign from positive to negative by increasing $Y$. The procedure is depicted in the inset of Fig.~\ref{fig:x1tbclevel}(a), where the energy gap as a function of $Y$ in the vicinity of $Y_0$ for $L = 14$ is shown as an example. As the model is on the boundary of the Haldane phase, the energy gap should go to zero in the thermodynamic limit for $Y < Y_c$. This is confirmed in Fig.~\ref{fig:x1tbclevel}(b), where the energy gap as a function of $1/L$ is plotted for $Y=0.1$. The data is fitted by a $4$-degree polynomial and it is seen that the extrapolated energy gap is indeed zero. We repeat this procedure for $L = 7, 8, \ldots, 20$ and determine each $Y_0$ with $\approx 10^{-9}$ precision, and then extrapolate the critical point as $Y_0(L) = Y_c + a L^{-2} + b L^{-4} + \ldots$. As shown in the main plot of Fig.~\ref{fig:x1tbclevel}, the extrapolated $Y_c$ is $0.3506694(3)$. The error is estimated by changing the degree of the polynomial.

\begin{figure}
  \centering
    \includegraphics[width=0.48\textwidth]{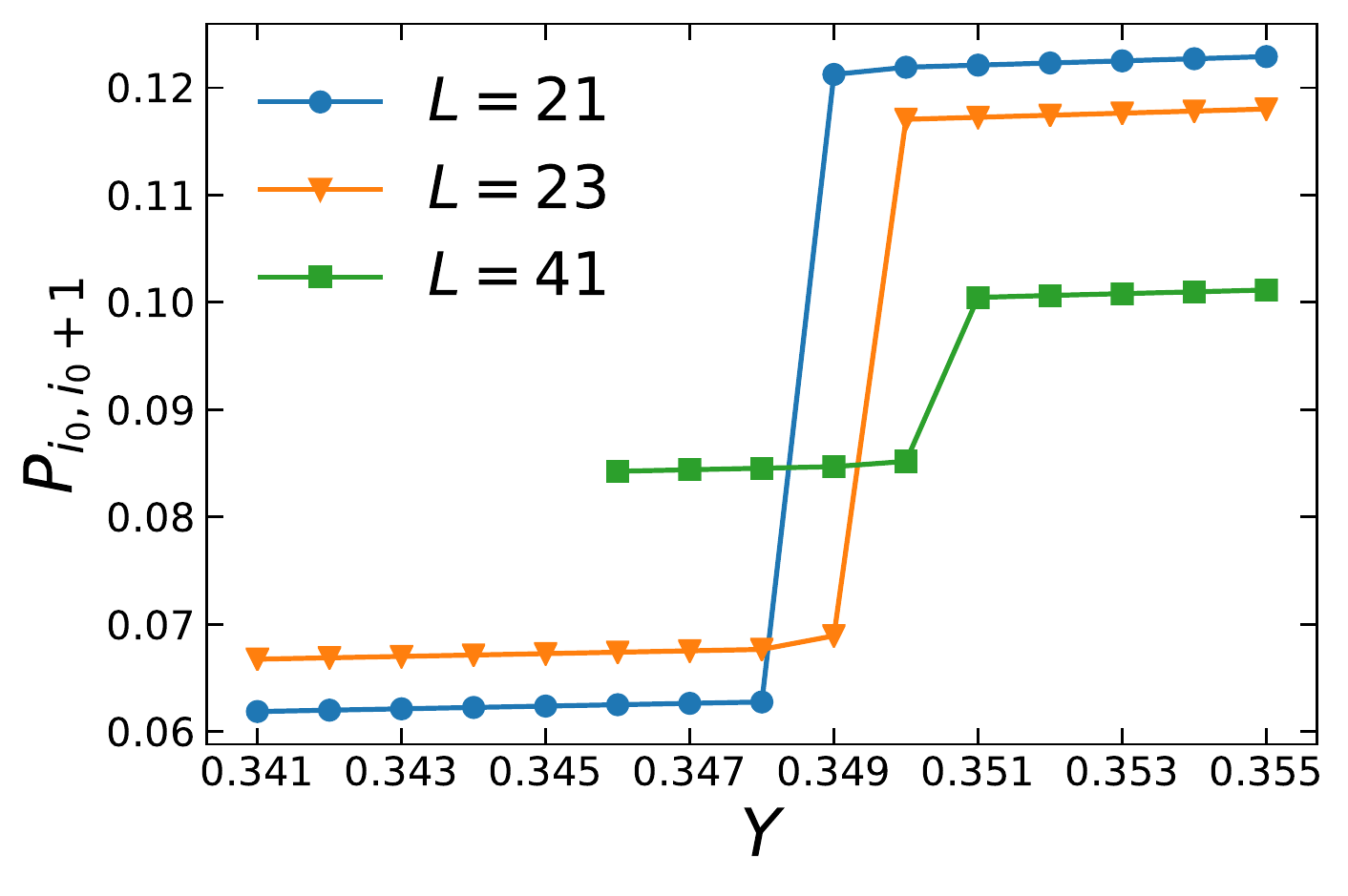}
    \caption{\label{fig:xm1oddlpermutation}DMRG calculation of the permutation operator $P_{i_0,i_0+1}[i_0 = (L+1)/2]$ as a function of $Y$ for $L = 21, 23, 41$ with PBCs. The discontinuity is between $0.348$ and $0.349$, $0.349$ and $0.350$, $0.350$ and $0.351$ for $L = 21, 23, 41$, respectively, consistent with $Y_c$ obtained in Fig.~\ref{fig:x1tbclevel}. }
  \end{figure} 
  
The level crossing also exists for $X = -1$, PBC and an odd number of sites. In this case, the total parity $P = -1$ for small $Y$ because there is an odd number of singlets in the ground state. The total parity is still $+1$ in the large-$D$ phase. We can calculate the energy difference between the ground-state energy in the sector $M = 0, k = 0, P = 1$, $E_{0, M = 0, k = 0, P = 1}$, and that in the sector $M = 0, k = 0, P = -1$, $E_{0, M = 0, k = 0, P = -1}$ and locate the position of level crossing. The values of $Y_0$ are exactly the same as those in Fig.~\ref{fig:x1tbclevel} for odd number of sites. Because the Hamiltonian with $X = -1$, PBCs and odd $L$ can be transformed to the one with $X = 1$, TBCs and odd $L$ just by rotating the spins on even or odd sites by an angle $\pi$. In Ref.~\cite{Langari_2013}, the same method is used but $Y_c$ is extrapolated with a power law and $Y_c = 0.347$ is obtained, different from our extrapolation. We check the result with the permutation operator $P_{i,j} = \vec{S}_i \cdot \vec{S}_j + (\vec{S}_i \cdot \vec{S}_j)^2 - \mathbf{1}$ proposed in Ref.~\cite{Langari_2013} for the odd-$L$ ring. The discontinuity in the permutation operator signals a phase transition changing parity. By using DMRG for $L = 41$ with PBCs, we show in Fig.~\ref{fig:xm1oddlpermutation} that the discontinuity is located between $Y = 0.350$ and $Y = 0.351$, consistent with our extrapolation. We also check for $L = 21$ and $L = 23$, and find that the discontinuity is between $Y = 0.348$ and $Y = 0.349$, and between $Y = 0.349$ and $Y = 0.350$, respectively. So the power-law extrapolation underestimates the phase-transition point. The sudden jump in the permutation operator shrinks as we increase the system size. It is expected that the discontinuity disappears in the thermodynamic limit, because the charge representation with $S = 1$ is always gapless for $Y < Y_c$ where the energy levels in the two parity sectors are degenerate.

\begin{figure}
  \centering
    \includegraphics[width=0.48\textwidth]{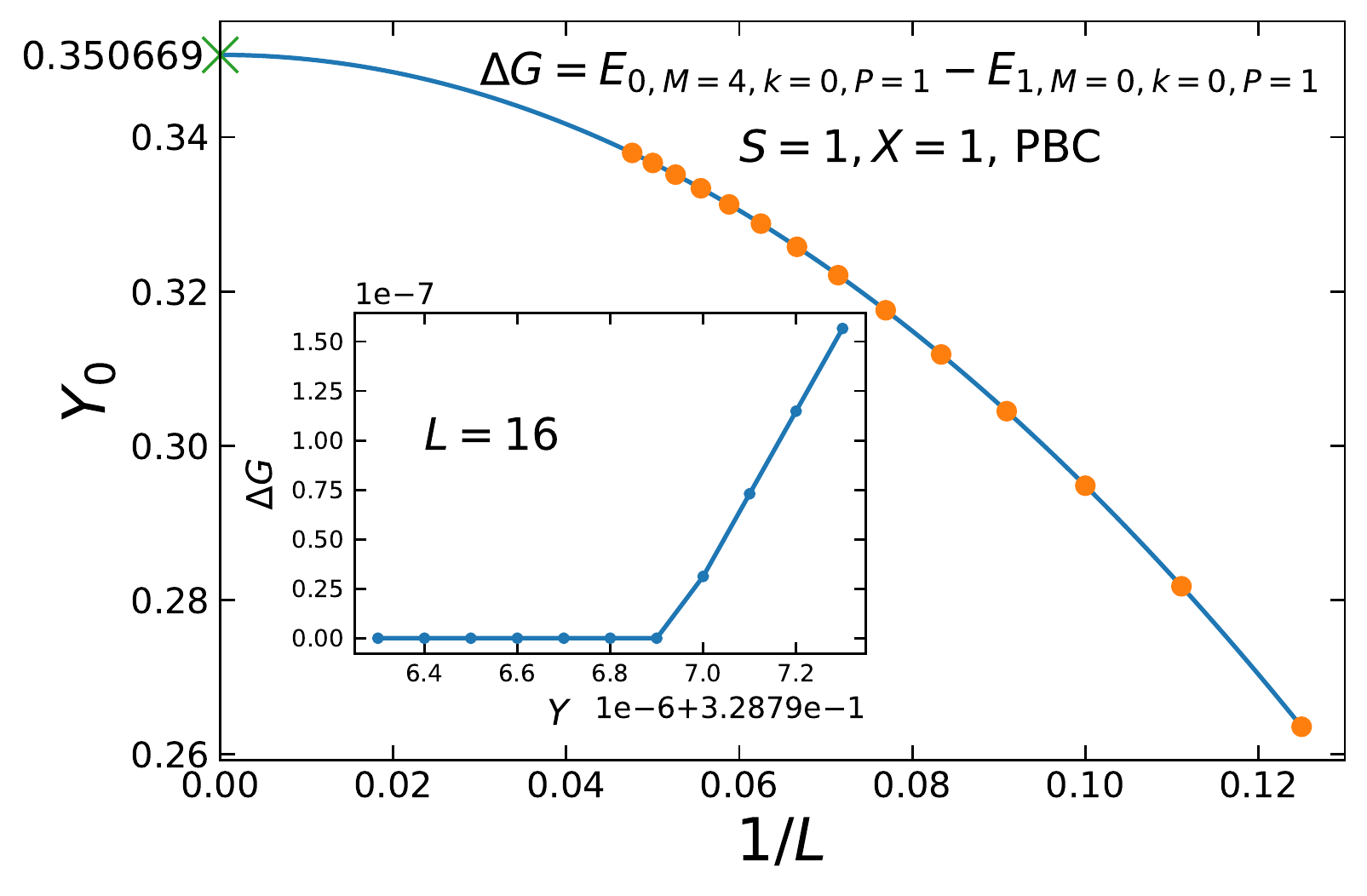}
    \caption{\label{fig:x1gapcross}Same as Fig. \ref{fig:x1tbclevel}. The results are for $S=1, X = 1$, PBCs. The finite size $Y_0$ up to $L = 21$ is found by locating the position where the energy difference $\Delta G$ between the ground-state energy of the sector $M=4, k=0, P=1$ and the first-excited state energy of the sector $M=0, k=0, P=1$ just closes. The inset shows $\Delta G$ versus $Y$ near $Y_0$ for $L = 16$. The extrapolated $Y_{c} = 0.35066928(2)$.
    }
  \end{figure}

Since the spin-$1$ truncation corresponds to the $J_z = 0$ limit of the $XXZ$ model with single-ion anisotropy, $Y_c$ is also the endpoint of the two BKT lines \cite{PhysRevB.67.104401}. The level crossing across the BKT critical line can also be applied here. Figure \ref{fig:x1gapcross} depicts our results for the level crossing between excitations classified by $(M= 4, k = 0, P = 1)$ and $(M = 0, k = 0, P = 1)$. The inset shows the procedure to locate $Y_0$ for $L = 16$. As expected, the two levels are exactly degenerate for $Y < Y_0$ because the system is on a BKT line. Again, this level crossing only happens for even $L$ if $X = -1$, but exists for all $L$ if $X = 1$. The extrapolated value for $Y_c$ is $0.35066928(2)$ for this method, consistent with the TBC method up to the seventh decimal place. Finally, we see that $Y_0$ at finite size from the TBC method is much closer to its thermodynamic value than this method for the same $L$. In principle, the operator content of BKT transitions with PBC can be related to the $k = 1$ SU(2) Wess-Zumino-Witten model by applying TBC \cite{doi:10.1143/JPSJ.66.3379}, where level crossings between lower excitations for finite-size systems can be used to locate a $Y_0$ value that is closer to the thermodynamic value. Our results show that the extrapolation procedure is very stable, and we will just apply the method described in Fig.~\ref{fig:x1gapcross} for spin-$2$ truncation and above.
  
\begin{figure}
  \centering
    \includegraphics[width=0.48\textwidth]{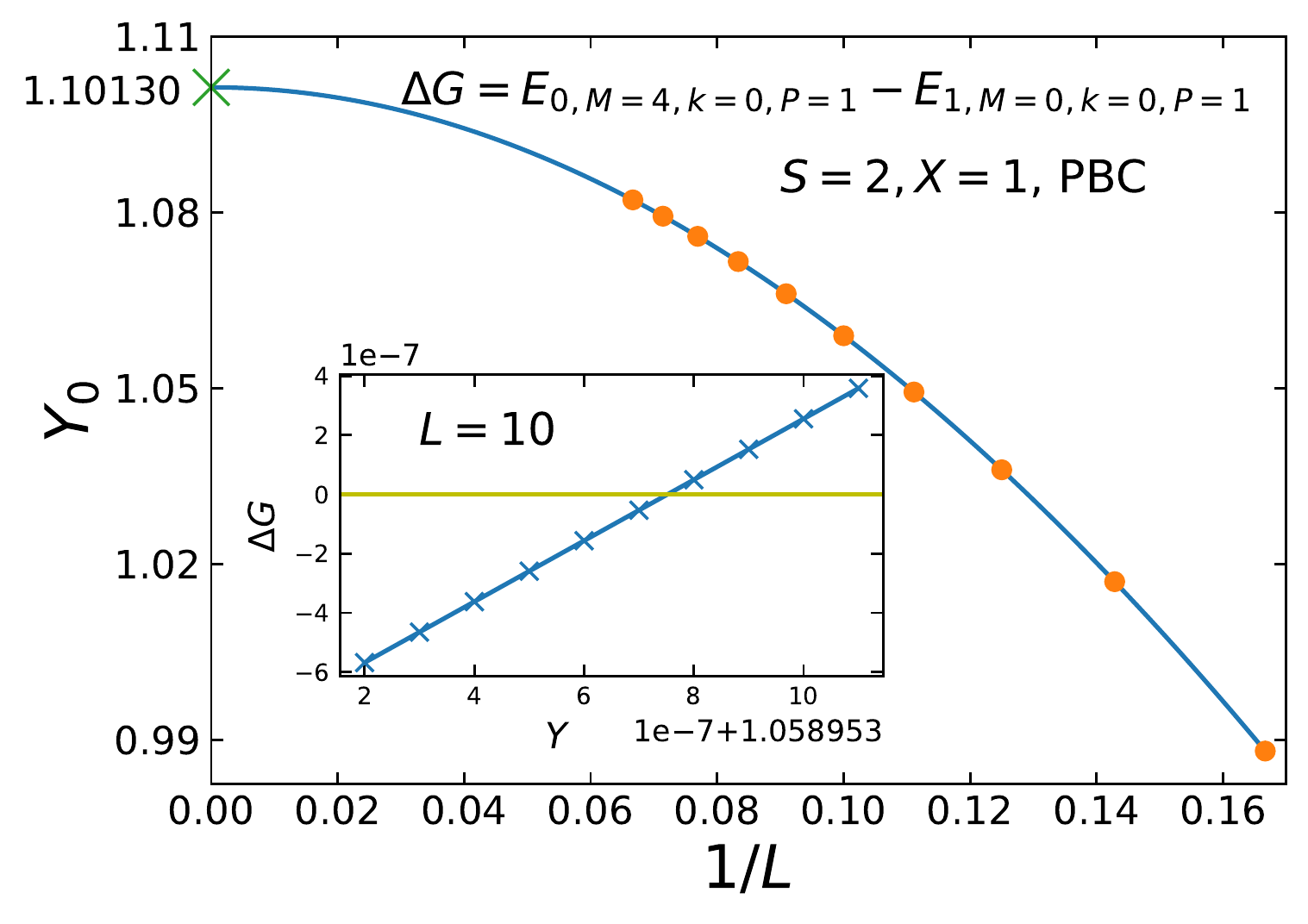}
    \caption{\label{fig:x2gapcross}Same as Fig. \ref{fig:x1gapcross}, but for $S = 2$. The system size used is up to $L = 15$. The extrapolated $Y_c= 1.101304(6)$. The inset shows the level crossing near $Y_0$ for $L = 10$.
    }
  \end{figure}
\begin{figure}
 \centering
\includegraphics[width=0.48\textwidth]{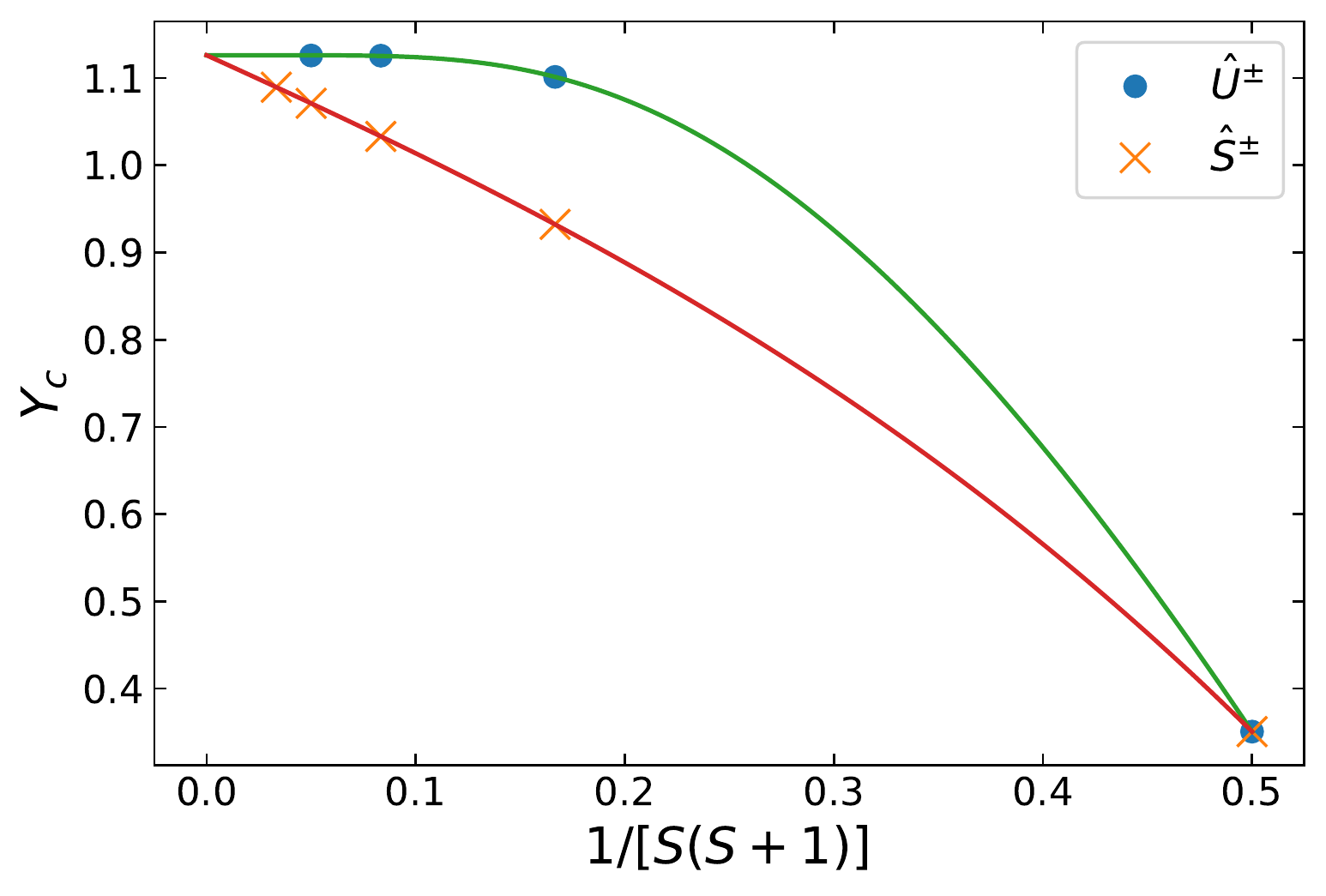}
\caption{\label{fig:ycvstrunc}The dependence of the BKT critical point $Y_c$ on spin truncation $S$. The solid line on solid circles is a curve fit with exponential convergence function of $S$. The solid line on cross symbols is a curve fit with a polynomial function of $1/\left[S(S+1)\right]$. The extrapolated $Y_{c} = 1.126188(13)$ and $1.12614(8)$ respectively.}
\end{figure}

Figure \ref{fig:x2gapcross} shows the extrapolation procedure for $S = 2$. In contrast to $S = 1$, there is no exact degeneracy for $Y < Y_0(L)$ and it is a true level crossing, as shown in the inset of Fig.~\ref{fig:x2gapcross}. The extrapolated value for $Y_c$ is $1.101304(6)$. This true level crossing persists for all $S \ge 2$ truncations, which means that the phase transitions really go across the BKT critical lines. In addition, we also calculate the transition points for the spin ladder operators $\hat{U}^{\pm} \rightarrow \hat{S}^{\pm} / \sqrt{S(S+1)}$. In Table~\ref{tableYcSs}, we summarize the transition points for $S = 1,2,3,4,5$ for both $\hat{U}^{\pm}$ and $\hat{S}^{\pm}$ operators.
\begin{table}  
    \begin{tabular} {p{2.5cm}  p{2.5cm}  p{2.5cm}}
    \hline
    \hline
        &$\hat{U}^{\pm}$ &$\hat{S}^{\pm}$  \\  \hline
    $S = 1$ &0.35066928(2) &0.35066928(2)  \\  \hline
    $S = 2$ &1.101304(6) &0.932201(4)   \\  \hline
    $S = 3$ &1.125614(17) &1.03308(3)   \\  \hline
    $S = 4$ &1.125898(19) &1.07103(2)   \\  \hline
    $S = 5$ & &1.08952(3)   \\  \hline
    $S = \infty$ &1.126188(13) &1.12614(8)   \\  \hline \hline
    \end{tabular}  
    \caption{\label{tableYcSs}Values of phase-transition points $Y_c$ for different $S$. Results are obtained by LS. }
\end{table}
The maximal $L$ in the extrapolation procedure is $13, 11, 10$ for $S = 3, 4, 5$, respectively. For $S = 5$, we only do the calculations with $\hat{S}^\pm$ operators. It is seen that $Y_c$ converges much faster with $S$ for $\hat{U}^{\pm}$ than it does for $\hat{S}^{\pm}$. We expect the convergence to be exponentially fast and fit $Y_c$ versus $S$ with $c + A \exp( - \alpha S)$ for $\hat{U}^{\pm}$ in Fig.~\ref{fig:ycvstrunc}. We find $\alpha = 3.4394(6)$ and the extrapolated value for $Y_c$ at infinite $S$ is $1.126188(13)$. Note that $\hat{S}^{\pm} / \sqrt{S(S+1)}$ differs from $\hat{U}^{\pm}$ in matrix elements that corresponds to raising (lowering) charges larger than $1$. Those matrix elements in $\hat{S}^{\pm} / \sqrt{S(S+1)}$ have a common factor $1/\sqrt{S(S+1)}$. We expect that $Y_c$ has polynomial scaling for $\hat{S}^{\pm}$ and fit the data with a polynomial function of $1/\left[S(S+1)\right]$ in Fig.~\ref{fig:ycvstrunc}. The extrapolated value for $Y_c$ is $1.12614(8)$ and agrees extremely well with that for $\hat{U}^\pm$ as expected. The exponential convergence behavior for $\hat{U}^{\pm}$ would help save atoms or qubits in the quantum simulation.

%%%%%%%%%%%%%%%%%%%%%%%%%%%%%%%%%%%%%%%%%%%%%%%%%%%%%%%%%%%%%%%%%%%%%
\subsection{Gap scaling} \label{subsec:resultsgaps}
%%%%%%%%%%%%%%%%%%%%%%%%%%%%%%%%%%%%%%%%%%%%%%%%%%%%%%%%%%%%%%%%%%%%%
\begin{figure}
  \centering
    \includegraphics[width=0.48\textwidth]{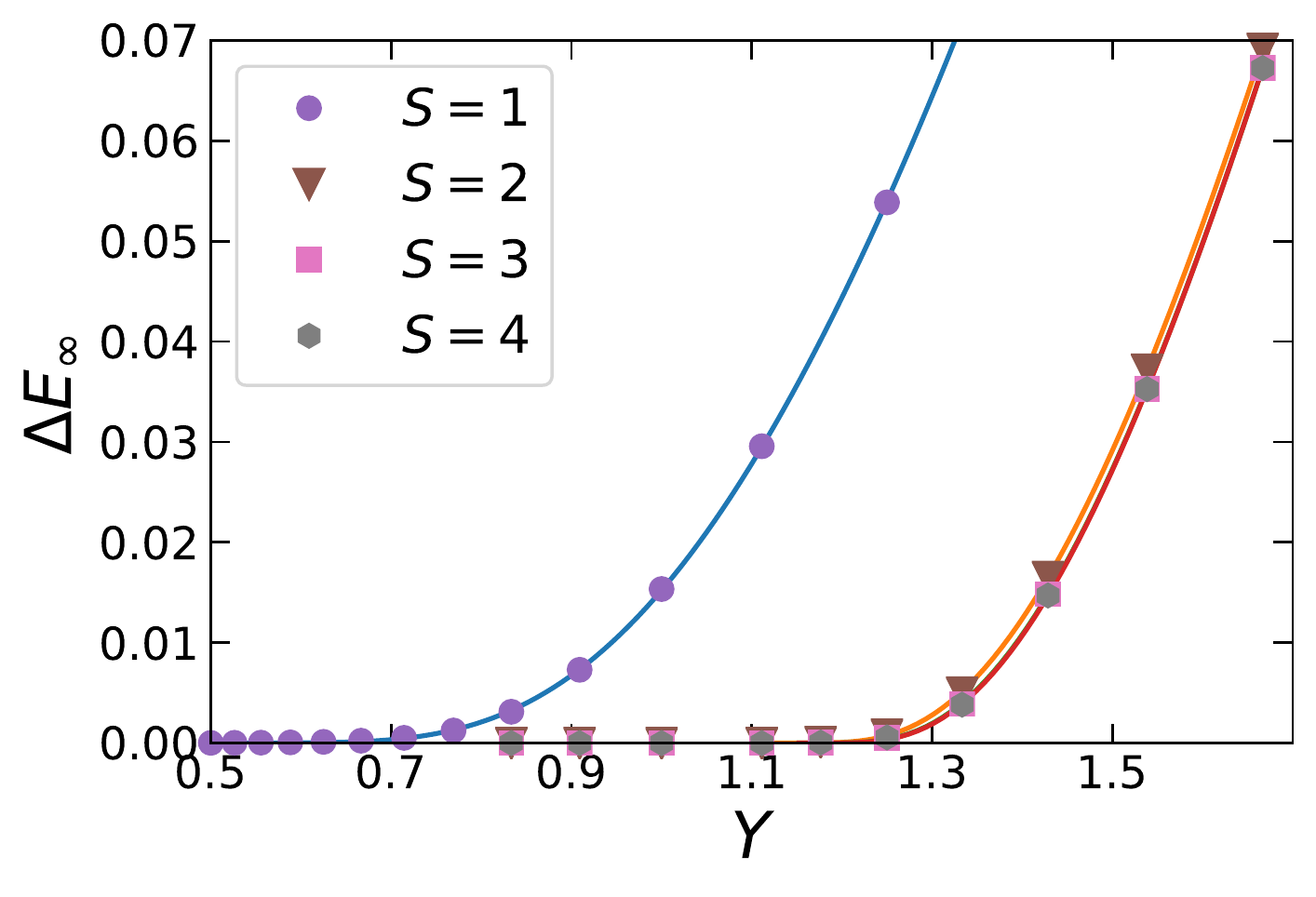}
    \caption{\label{fig:gaps1-4inf}Extrapolated energy gaps of O(2) Hamiltonian in the charge representation for spin truncations $S = 1, 2, 3, 4$ in the thermodynamic limit, as a function of $Y$. The solid lines on the symbols are curve fits with $A \sqrt{Y-Y_c} \exp\left[-b/(Y-Y_c)\right]$ for $S = 1$ and $A \exp(-b/\sqrt{Y-Y_c})$ for $S \ge 2$. }
  \end{figure}
  
\begin{figure}
  \centering
    \includegraphics[width=0.48\textwidth]{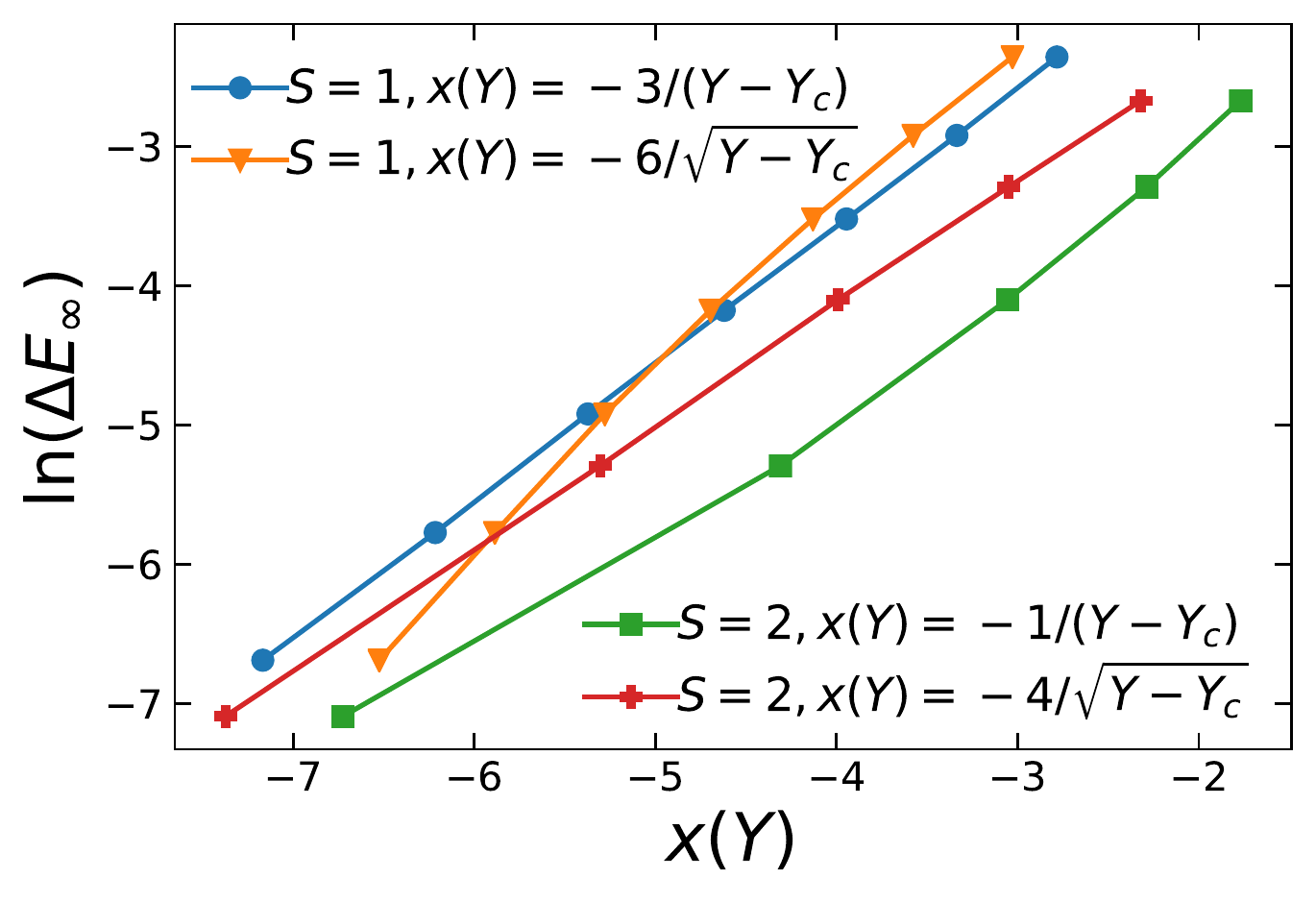}
    \caption{\label{fig:gaps1-2infvsxY}The logarithm of the extrapolated energy gaps of O(2) Hamiltonian in the charge representation for spin truncations $S = 1, 2$ as a function of $x(Y)$. The definition of $x(Y)$ is described in the legend. The curves are shifted for a better view. The linearity of blue circles and red pluses confirms the different essential singularities for $S = 1$ and $S = 2$.}
  \end{figure}
  
We have shown that the results from LS are extremely accurate. In this section, we use the scaling of the energy gap for the first-excited state to detect the infinite-order phase transitions in the charge representation, and compare it with the LS method. For all $S$ truncations, the ground state is inside the charge-zero sector, and the first-excited state is inside the charge-one sector. Figure \ref{fig:gaps1-4inf} shows the extrapolated energy gaps in the thermodynamic limit, $\Delta E_{\infty}$, as a function of $Y$ for different spin truncations $S = 1, 2, 3, 4$. The extrapolation procedure uses gaps of systems with up to $1024, 768, 512, 384$ sites for $S = 1, 2, 3, 4$ respectively and fit the data with high degree polynomials. We see similar behavior as the magnetic susceptibility shown in Fig.~\ref{fig:msus1-4inf}. The energy gap for $S = 1$ is very different from those for $S \ge 2$. It converges very fast with the spin truncation $S$ and almost already converges at $S = 2$. The data points for spin-$3$ and spin-$4$ truncation have differences that are not visible by eye and stay on top of each other in the plot of Fig.~\ref{fig:gaps1-4inf}. The energy gap vanishes at small $Y$ for all $S$, indicating a gapped-to-gapless phase transition. These results are consistent with exponential convergence of the phase-transition points obtained by LS. As for a small enough distance to the phase-transition point $\Delta Y = Y-Y_c$, $\exp\left(-b/\Delta Y\right) < \exp\left(-b'/\sqrt{\Delta Y}\right)$, the energy gap for $S \ge 2$ is much larger than that for $S = 1$ for the same $\Delta Y$. In other words, the energy gap for $S = 1$ stays extremely small for a large range of $Y > Y_c$, which makes it difficult to determine the place where the gap closes. We can take an initial estimate for the point where the gap closes by looking at where the center of the marker symbol approaches zero in Fig.~\ref{fig:gaps1-4inf}. For $S = 1$, $Y_c < 0.6$, while for $S \ge 2$, $Y_c < 1.17$.

\begin{figure}
  \centering
    \includegraphics[width=0.48\textwidth]{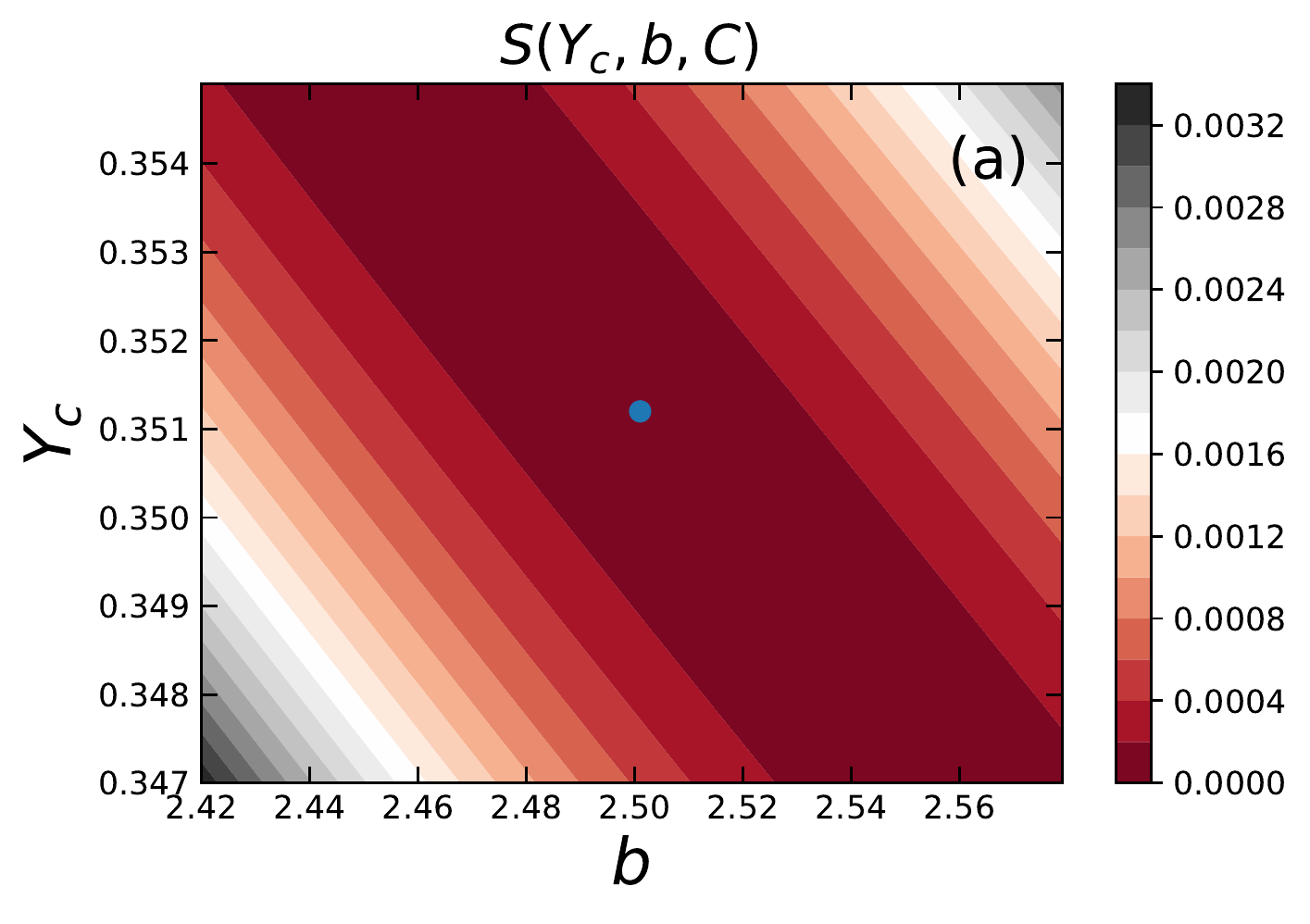}
    \includegraphics[width=0.48\textwidth]{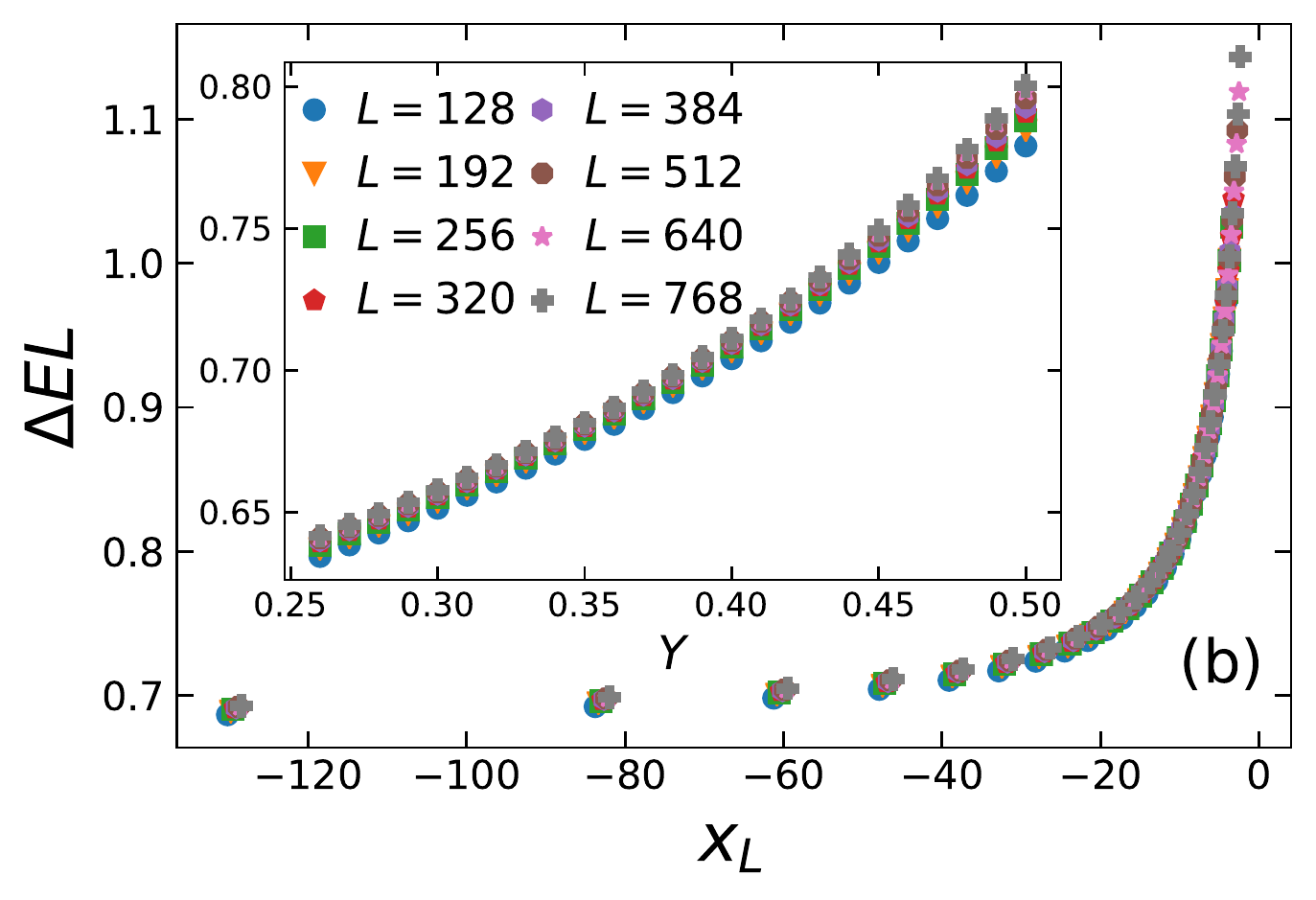}
    \caption{\label{fig:s1gapscaling}(a) Contour plot of the sum of squared residuals $S(Y_c,b,C)$ for $S = 1$. $S(Y_c,b,C)$ is minimized at $Y_c = 0.3512, b = 2.501, C = \infty$. (b) The best data collapse of $\Delta E L$ vs $x_L = \ln(L) - b/(Y-Y_c) + \ln(Y-Y_c)/2$ for $S = 1$. The inset shows $\Delta E L$ as a function of $Y$. 
    }
  \end{figure}
Now we fit the extrapolated energy gap with $\Delta E = A\sqrt{Y-Y_c} \exp\left[-b/(Y-Y_c)\right]$ for $S = 1$ and $\Delta E = A \exp\left(-b/\sqrt{Y-Y_c}\right)$ for $S \ge 2$. As the essential singularity results in a tiny energy gap near the critical point, the extrapolated data need high precision in the curve fit. Our DMRG data have small enough error ($\approx 10^{-8}$) thus the main error comes from the extrapolation procedure. The results for $S = 1,2,3,4$ are summarized in Table~\ref{tableYcSsInfgap}.
\begin{table}[ht]
%\scriptsize %\footnotesize
\centering     
\begin{tabular} {p{2.5cm}  p{2.5cm}  p{2.5cm}}
\hline
\hline
        &$Y_c$ &$b$  \\  \hline
$S = 1$ &0.368(7) &2.45(4)  \\  \hline
$S = 2$ &1.120(5) &3.21(6)   \\  \hline
$S = 3$ &1.144(6) &3.08(7)   \\  \hline
$S = 4$ &1.147(7) &3.06(8)   \\  \hline \hline
\end{tabular}  
\caption{\label{tableYcSsInfgap}Values of phase-transition points $Y_c$ and $b$ for different $S$ with $\hat{U}^{\pm}$ operators. Results are obtained by fitting the extrapolated energy gaps in Fig.~\ref{fig:gaps1-4inf} with $\Delta E = A\sqrt{Y-Y_c} \exp\left[-b/(Y-Y_c)\right]$ for $S = 1$ and $\Delta E = A \exp\left(-b/\sqrt{Y-Y_c}\right)$ for $S \ge 2$.}
\end{table}
They are all close to the results from LS and only differ in the second decimal place, which means that our polynomial extrapolations are accurate. In particular, the result for $S = 1$ has about $5\%$ relative error, while the results for $S \ge 2$ have less than $2\%$ error. All the results are larger than those from LS because the essential singularity has corrections away from the phase-transition point in the gapped phase. If we use the BKT formula of energy gap for $S = 1$, we obtain $Y_c = 0.514(8)$, far from the result from LS. We can also discriminate the two essential singularities by plotting the logarithm of the extrapolated energy gap as described in Fig.~\ref{fig:gaps1-2infvsxY}. We see that $\ln(\Delta E_{\infty})$ is more linear when plotted versus $1/(Y-Y_c)$ than versus $1/\sqrt{Y-Y_c}$ for $S = 1$, while it is more linear as a function of $1/\sqrt{Y-Y_c}$ for $S = 2$.

Another observation is that the extrapolated energy gaps become negative near $Y_c$ (not shown here): around $0.365$ for $S = 1$, around $1.115$ for $S = 2$, and around $1.130$ for $S = 3$. The numbers are even closer to Table~\ref{tableYcSs} than are those in Table~\ref{tableYcSsInfgap}. Obviously, the negative extrapolated energy gaps are not correct. The reason is that there should be logarithmic corrections in the scaling of energy gaps in the gapless phase. The polynomial fitting is not enough to accurately capture the finite-size scaling of the energy gap. However, the smallness of these negative numbers (of order of $10^{-6}$ or less) indicates that the logarithmic corrections are small, which explains why we obtain good results from the polynomial extrapolation of the energy gap. In the following, we apply the ansatz of the scaling of the energy gap at finite size and show that the logarithmic corrections near $Y_c$ are indeed highly suppressed, at least for OBC considered here.
  
\begin{figure}
  \centering
    \includegraphics[width=0.48\textwidth]{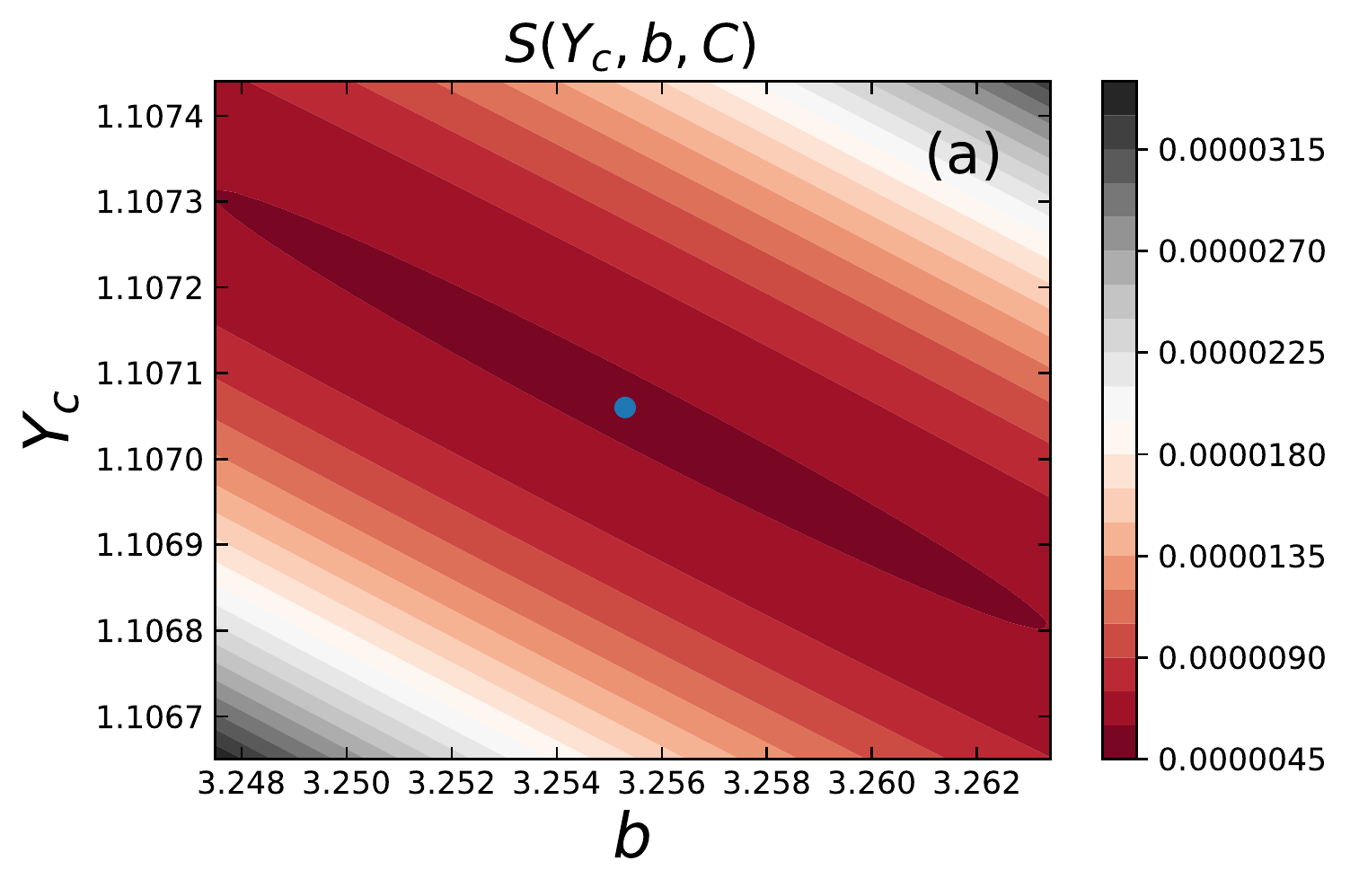}
    \includegraphics[width=0.48\textwidth]{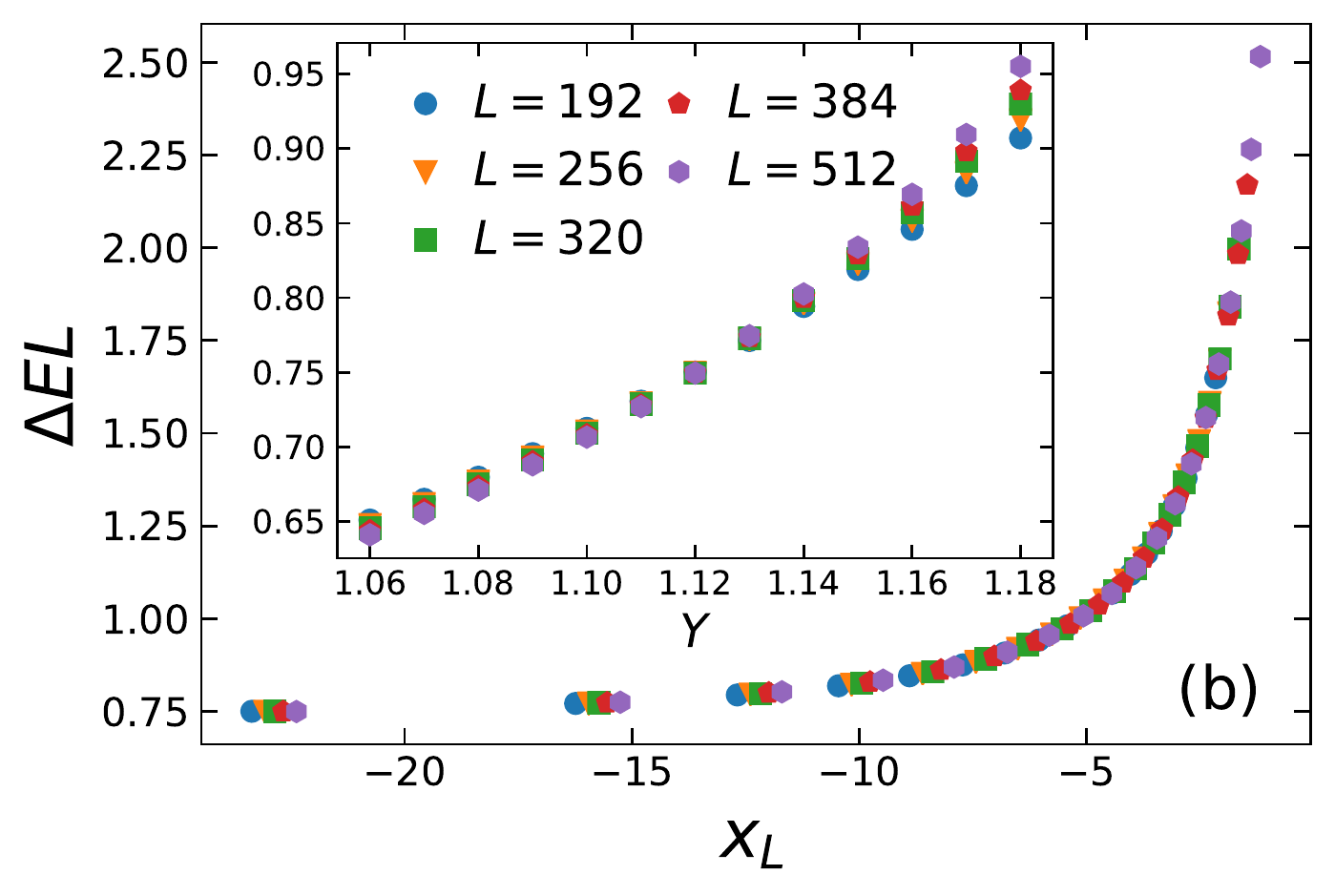}
    \caption{\label{fig:s2gapscaling}The same as Fig.~\ref{fig:s1gapscaling}, but for $S = 2$ and $x_L = \ln(L) - b/\sqrt{Y-Y_c}$. The sum of squared residuals is minimized at $Y_c = 1.10706, b = 3.2553, C = \infty$. 
    }
  \end{figure}

The above method will fail in a system with a very large $b \gg 1$, where the gap may be below the machine precision even though $Y-Y_c$ is not so small, and the extrapolation will be highly unreliable. We apply a more stable method using the ansatz of the scaling of the energy gap in Eq. \eqref{eq:gapscaling}. This method does not require extrapolation of the energy gap, and is more accurate. The correction term is taken to be $g(L) = 1/(2\ln L+C)$. Following \cite{PhysRevB.84.115135, PhysRevA.87.043606, PhysRevB.91.165136}, we first calculate the energy gap for different values of $Y$ and different system sizes. We adjust $Y_c, b, C$, calculate the rescaled gap $\Delta E_s$ and $x_L = \ln{L} - b/(Y-Y_c) + \ln(Y-Y_c)/2$ for $S = 1$ and $x_L = \ln{L} - b/\sqrt{Y-Y_c}$ for $S \ge 2$, fit $\Delta E_s$ versus $x_L$ with an arbitrary high degree polynomial, and find the best $Y_c, b, C$ that minimize the sum of squared residuals $S(Y_c, b, C)$. In practice, we choose the data set that is robust to adding or removing data. The results for $S = 1$ are depicted in Fig.~\ref{fig:s1gapscaling}(a), the sum of squared residuals is minimized at $Y_c = 0.3512(10), b = 2.501(13)$ using data with $L \ge 320$, and $C$ is arbitrarily large as expected for Gaussian points. The error is estimated by adding or removing nearby data. The result for $Y_c$ is much closer to Table~\ref{tableYcSs} than that from extrapolated energy gaps. The perfect data collapse of $\Delta E L$ versus $x_L$ is seen in Fig.~\ref{fig:s1gapscaling}(b), where all the rescaled energy gaps for $L = 128, 192, 256, 320, 384, 512, 640, 768$ collapse onto a single smooth curve. 

For $S = 2$, the result is much more stable, with smaller uncertainty. As shown in Fig.~\ref{fig:s2gapscaling}(a), the sum of squared residuals is minimized at $Y_c = 1.10706(7), b = 3.2553(21)$, and $C$ is again arbitrarily large. The best data collapse is depicted in Fig.~\ref{fig:s2gapscaling}(b). Comparing Fig.~\ref{fig:s1gapscaling}(a) with Fig.~\ref{fig:s2gapscaling}(a), it is seen that the structure of the contour map of $S(Y_c, b, C)$ for $S = 1$ is very different from that for $S = 2$. For $S = 2$, the contours form a clear ellipse in a very narrow region of $(Y_c, b)$, while it is difficult to see an ellipse for $S = 1$, indicating that the gradient of $S(Y_c, b, C)$ in one direction is very small. We also consider adding a higher-order correction term $A / \ln^2(L)$, and find that $Y_c = 1.1033(3), b = 3.334(6), A = -0.295(14)$. By adding this correction term, $Y_c$ is closer to the result obtained by LS $1.101304(6)$. Adding the $A / \ln^2(L)$ correction term only changes the third decimal place for $Y_c$, but minimization of $S(Y_c, b, C, A)$ in four parameter space takes much more time. We only consider three parameters for other cases. The results from the ansatz of the scaling of the energy gap is summarized in Table~\ref{tableYcSsgapscaling}.
\begin{table}[ht]
\centering     
\begin{tabular} {p{2.5cm}  p{2.5cm}  p{2.5cm}}
\hline
\hline
        &$\hat{U}^{\pm}$ &$\hat{S}^{\pm}$  \\  \hline
$S = 1$ &\makecell[l]{0.3512(10) \\ 2.501(13)} &\makecell[l]{0.3512(10) \\ 2.501(13)}  \\  \hline
$S = 2$ &\makecell[l]{1.10706(7) \\ 3.2553(21)} &\makecell[l]{0.93978(15) \\ 3.647(4)}   \\  \hline
$S = 3$ &\makecell[l]{1.13191(14) \\ 3.110(5)} &\makecell[l]{1.03933(8) \\ 3.367(2)}   \\  \hline
$S = 4$ &\makecell[l]{1.13213(16) \\ 3.117(5)} &\makecell[l]{1.0767(1) \\ 3.281(3)}   \\  \hline
$S = 5$ & &\makecell[l]{1.0948(3) \\ 3.25(1)}   \\  \hline \hline
\end{tabular}  
\caption{\label{tableYcSsgapscaling}Values of phase-transition points $Y_c$ (first line) and $b$ (second line) for different $S$. Results are obtained by the gap scaling ansatz with $g(L) = 1/\left[2\ln(L)+C\right]$. $C = \infty$ for all cases. Including a higher-order correction $A/\ln^2(L)$ can further improve the results, e.g., $Y_c = 1.1033(3)$ for $S = 2$. }
\end{table}
Compared with Table~\ref{tableYcSs}, the difference in $Y_c$ from the gap scaling ansatz is order of $10^{-3}$, less than $0.5\%$.

Note that $C = \infty$ for all the cases, which means that the logarithmic corrections are highly suppressed near the phase transition. This also happens for the one-dimensional Bose Hubbard model with OBC \cite{PhysRevA.87.043606}, and the spin-$3/2$ $XXZ$ chain with OBC \cite{PhysRevB.91.165136}. In Refs.~\cite{PhysRevA.87.043606, PhysRevB.91.165136}, PBCs are also considered and $C$ is finite. These models, including ours, are all bosonic and have a global U(1) symmetry, and it seems that OBC suppresses the first-order logarithmic corrections near the phase-transition point. For fermionic systems with OBC, $C$ is also finite \cite{PhysRevB.91.165136}. If a phase transition goes across a BKT critical line, near the BKT line, the finite-size effects of the scaling dimensions related to the excitation in the critical phase behave differently from that in the gapped phase. This effect appears in the energy gap as a crossing point of the rescaled gap $\Delta E L [ 1 + g(L)]$ near but larger than the phase-transition point, as shown in the inset of Fig.~\ref{fig:s2gapscaling}(b) for $S = 2$. In the procedure of finding the best data collapse, the variable $Y$ is first rewritten as $-b/\sqrt{Y-Y_c}$ and then shifted by $\ln L$, the single crossing point separates into multiple points that the universal function must go through, which largely suppress the uncertainty in the optimization procedure and pull the value of optimized $Y_c$ to the gapped side. For the infinite-order Gaussian transition to a BKT critical line ($S = 1$), the rescaled energy gap as a function of $Y$ would just approach to the thermodynamic value from below without a crossing point near $Y_c$. This behavior is presented in the inset Fig.~\ref{fig:s1gapscaling}(b), where the gapped side ($Y > Y_c$) is similar to the finite-order Gaussian transition \cite{PhysRevB.99.134408, PhysRevB.102.064414}. In this case, on one hand, there is still a point, where the rescaled energy gaps have minimal distances, that plays the same role as the crossing point in BKT transition. On the other hand, all the values of $\Delta E L$ are below the true collapsed line in the thermodynamic limit, so is the best fit data collapse using finite-size energy gaps. Therefore, $Y_c$ should be smaller to compensate this difference. Overall, we obtain a result that has the smallest discrepancy from that by LS. 

Finally, we believe that the discrepancy between Tables \ref{tableYcSsgapscaling} and \ref{tableYcSs} is from higher-order corrections for the energy gap near the critical point. One piece of evidence is that the result for $S = 2$ becomes closer to that from LS by adding a higher-order correction term $A/\ln^2(L)$. It is expected to have more accurate result by considering more correction terms. However, the results only have an order of $10^{-3}$ discrepancy from LS by considering only the leading correction term. This is the advantage of this method in locating infinite-order phase transitions.

%%%%%%%%%%%%%%%%%%%%%%%%%%%%%%%%%%%%%%%%%%%%%%%%%%%%%%%%%%%%%%%%%%%%%
\subsection{Correlation-function exponent} \label{subsec:correlationexponent}
\begin{figure}
  \centering
    \includegraphics[width=0.48\textwidth]{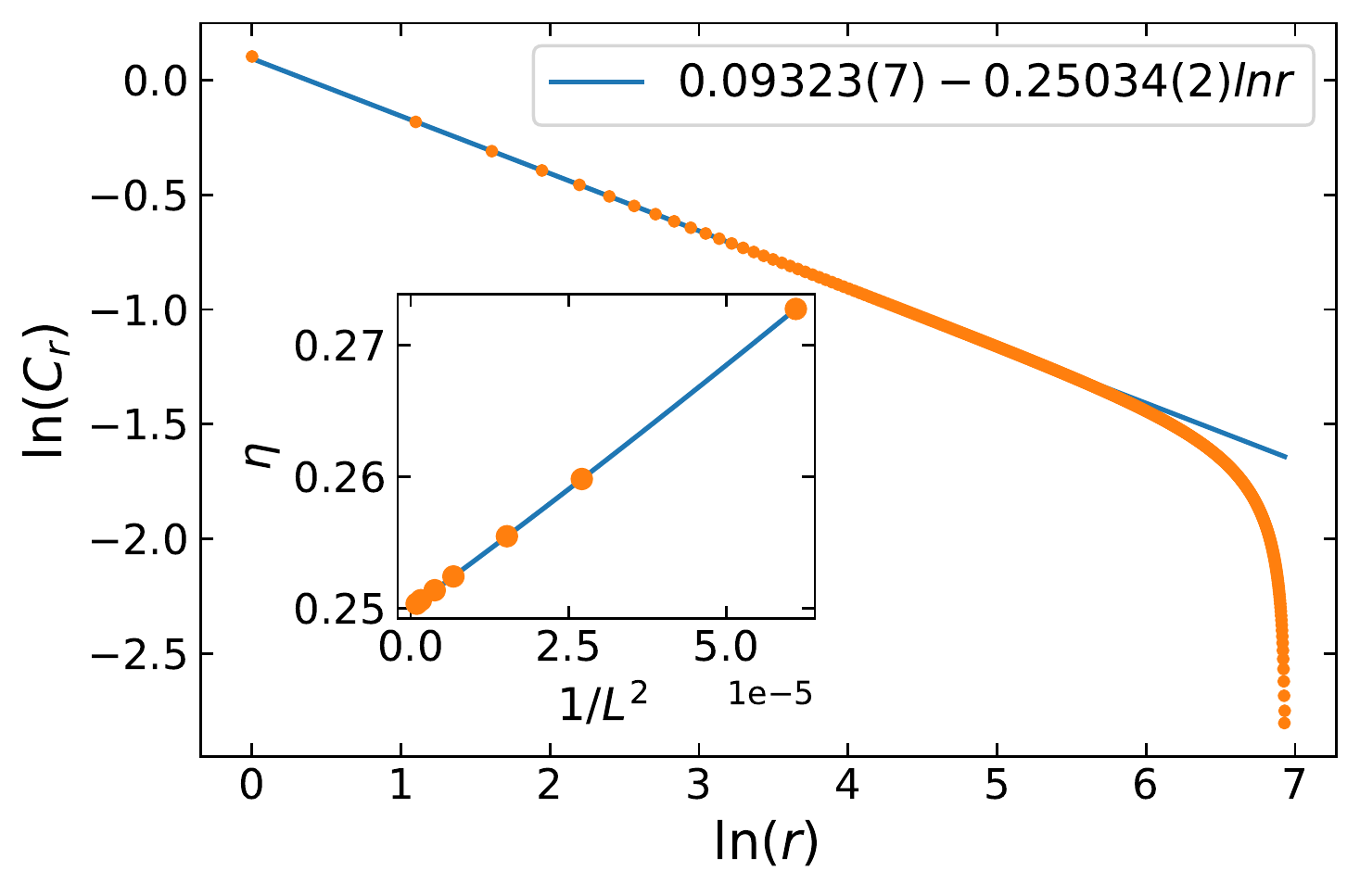}
    \caption{\label{fig:s1corr1024}Log-log plot of the correlation function $C_r$ as a function of $r$ ($r = 1, 3, 5, \ldots$) for $S = 1, Y = 0.35067$. The linear fit is performed with $r = 21, 23, \ldots, 39$. The inset shows the extrapolation of correlation exponent to $\eta_{\infty} = 0.24997(6)$.
    }
  \end{figure}

%%%%%%%%%%%%%%%%%%%%%%%%%%%%%%%%%%%%%%%%%%%%%%%%%%%%%%%%%%%%%%%%%%%%%
The multiplicative logarithmic corrections stemming from the marginal operators often stand in the way of calculating the critical exponents accurately. An advantage of $S = 1$ truncation is that the coupling constant of the marginal operators becomes zero at the infinite-order Gaussian transition point connecting the BKT critical lines, where the logarithmic corrections vanish with the same critical exponents as BKT. We can then extract the critical exponents accurately without going to very large system sizes. As an example, we calculate the correlation function
\begin{eqnarray}
\label{eq:correlator}
C_r = \langle U^+_{L/2-(r-1)/2} U^-_{L/2+(r-1)/2+1}\rangle \sim \frac{1}{r^{\eta}}
\end{eqnarray}
for $S = 1$. Figure \ref{fig:s1corr1024} shows the plot of $\ln(C_r)$ versus $\ln(r)$ for $L = 1024$. Far from the boundary, the plot is perfectly linear, and a linear fit for data with $r = 21, 23, \ldots, 39$ gives the correlation-function exponent $\eta = 0.25034(2)$, close to the expected value for BKT transitions $1/4$. The same procedure is performed for $L = 128, 192, \ldots, 768$ and the results are presented in the inset of Fig.~\ref{fig:s1corr1024}. A polynomial fit of $\eta(L)$ versus $1/L^2$ gives the extrapolated $\eta = 0.24997(6)$. The accurate determination of the correlation-function exponent from just linear fits in turn confirms that there are no multiplicative logarithmic corrections to the correlation function at the quantum phase transition for $S = 1$ truncation.
%%%%%%%%%%%%%%%%%%%%%%%%%%%%%%%%%%%%%%%%%%%%%%%%%%%%%%%%%%
\section{Conclusions} \label{sec:conclusion}
%%%%%%%%%%%%%%%%%%%%%%%%%%%%%%%%%%%%%%%%%%%%%%%%%%%%%%%%%%
In the context of compact sQED, the O(2) model is the zero-gauge-coupling limit where only matter field interaction exists. By Fourier transforming the compact variables, a dual representation called the charge representation can be obtained where the discrete variables have the physical meaning of electric charge quantum numbers. The quantum Hamiltonian can be obtained by taking the time continuum limit. In ($1+1$) dimensions, the O(2) model has a nontrivial BKT phase transition that is important to explain fundamental phenomena of condensed-matter physics and gauge theories. However, due to the essential singularity of the correlation length resulting in an exponentially small energy gap, and the logarithmic corrections stemming from the marginal operator, it is difficult for classical computing to detect the BKT transition for both the path integral formulation and the quantum Hamiltonian. We expect that the accurate manipulation and measurement of atoms or qubits in the future would overcome this difficulty. Spin-$1$ models can be realized by a spin-$1/2$ two-legged ladder \cite{sompet2021realising}. A two-species Bose-Hubbard model is suitable for quantum simulating the charge representation with spin $S$ truncation \cite{PhysRevA.90.063603, PhysRevD.92.076003}, where large onsite interactions and a chemical potential are tuned so that there are $2S$ particles per site. Building these models allows us to study more intriguing dynamics in quenches from one phase to another as is done in Ref.~\cite{dhar2021dynamics}, which may also present interesting truncation effects.

To stimulate the efficient manipulation of an increasing number of atoms or qubits in the near future, it is important to figure out what truncations and system sizes are needed to study BKT transitions. In this paper, we discussed the truncation effects of the quantum phase transition in the charge representation. We found that there is always an infinite-order phase transition for any integer $S$ in the charge representation, but the $S = 1$ truncation is different from $S \ge 2$ truncations. There is a hidden SU(2) symmetry in the charge representation for $S=1$, where the phase transition is from a gapped phase into a BKT critical line. The transition point is an infinite-order Gaussian point described by the $k = 1$ SU(2) Wess-Zumino-Witten CFT. The same type of phase transition can be observed in the explicit SU(2) symmetric models such as the Hubbard model \cite{PhysRevB.60.7850} and the $J_1 - J_2$ antiferromagnetic Heisenberg spin-$1/2$ chain \cite{PhysRevB.25.4925}. The originally defined BKT transition in the O(2) model is observed in $S \ge 2$ truncations. The essential singularities are different and the correlation length diverges as $(Y-Y_c)^{-1/2}\exp\left[b/(Y-Y_c)\right]$ for $S = 1$ and $\exp(b/\sqrt{Y-Y_c})$ for $S \ge 2$. By applying the level spectroscopy (LS) method, we obtained the phase-transition point accurately and found that the phase-transition point converges exponentially with $S$ for the truncated $\hat{U}^{\pm} = \exp(\pm i \hat{\theta})$ operators, while it converges polynomially with $1/S$ for the spin ladder operators $\hat{S}^{\pm}$ that are often used in quantum link models.

As LS is accurate, our models are prime candidates to test other universal methods for detecting quantum phase transitions. Those methods only require calculating the low-energy states and no prior analysis of critical properties of the model is needed. In ($1+1$) dimensions, the powerful DMRG algorithm can make these methods efficient and accurate. One of them is to make use of the energy gap between the lowest two levels with OBC. We first extrapolated the energy gap to the thermodynamic limit, and fit the extrapolated values with $\Delta E \sim (Y-Y_c)^{1/2}\exp\left[-b/(Y-Y_c)\right]$ for $S = 1$ and $\Delta E \sim \exp(-b/\sqrt{Y-Y_c})$ for $S \ge 2$. The results have only order of $10^{-2}$ discrepancy with those from LS. We then used the ansatz for the scaling of the finite-size energy gap described in Eq. \eqref{eq:gapscaling}. By calculating the energy gaps for various values of $Y$ and $L$ near the phase-transition point in the gapped phase and minimizing the sum of squared residuals in the procedure of finding the best data collapse, we were able to locate the phase-transition points with discrepancy of order of only $10^{-3}$. Using the correct essential singularity behavior for the correlation length for $S=1$ truncation is crucial to obtain the accurate result. We also found that the logarithmic corrections in the finite-size energy gap is highly suppressed, which is also seen in the one-dimensional Bose Hubbard model \cite{PhysRevA.87.043606} and the spin-$3/2$ $XXZ$ chain \cite{PhysRevB.91.165136}. It is believed that it is the open boundary condition (OBC) that suppresses the logarithmic corrections in these bosonic models, while the fermionic Hubbard models have nonnegligible logarithmic corrections even with OBCs \cite{PhysRevB.91.165136}. A similar cancellation of logarithmic corrections in the $XXX$ spin-$1/2$ chain can be derived with a large edge magnetic field in the $x$ direction \cite{PhysRevB.62.5546}.

Finally, $S = 1$ truncation moves the BKT transition point to a Gaussian point where the logarithmic corrections vanishes but critical exponents $\delta, \eta$ stay the same. Thus we can measure the critical properties of BKT transitions without going to very large systems where the logarithmic corrections is not important. It is interesting if this phenomena can be seen in other models that have BKT transitions. In general, one may think about whether we can manipulate the truncation nontrivially to impose explicit SU(2) symmetry, in such a way that the BKT transition becomes infinite-order Gaussian. It is difficult to see the trivial truncations with a hidden SU(2) symmetry unless an accurate phase diagram is determined in advance as is done in this paper, but it is interesting to study in what kind of systems this can happen. These types of considerations can be explored in the design of minimal experimental implementations required for quantum simulations of given critical properties.

\vskip20pt
\begin{acknowledgments}
We thank G. Ortiz and J. Unmuth-Yockey for helpful discussions. This work was supported in part by the National Science Foundation (NSF) RAISE-TAQS under Award Number 1839153 (S.W.T.) and by the U.S. Department of Energy (DOE) under Award Number DE-SC0019139 (Y.M.). Computations were performed using the computer clusters and data storage resources of the HPCC, which were funded by grants from NSF (MRI-1429826) and NIH (1S10OD016290-01A1).
\end{acknowledgments}

\appendix 
\begin{figure}
  \centering
    \includegraphics[width=0.48\textwidth]{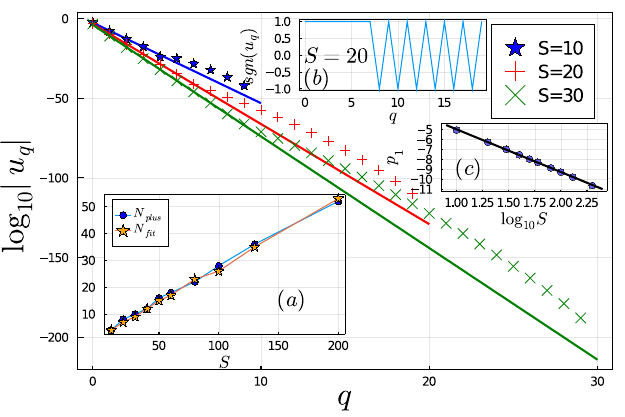}
    \caption{\label{fig:upmcoeffs}The logarithm of the magnitude of coeffients $u_q$ in Eq.~\eqref{eq:oprelation} as a function of the index $q$. The solid lines are linear fits of the first four, seven and nine points for $S = 10, 20$ and $30$ respectively. Inset $(a)$ shows the number of positive $u_q$, $N_{plus}$, and the number of data on the linear fits, $N_{fit}$, as a function of $S$. Inset $(b)$ shows the signs of $u_q$ as a function of $q$. Inset $(c)$ shows the slopes of the linear fits, $p_1$, as a function of $\log_{10}S$.
    }
  \end{figure}
  
\section{The sine-Gordon theory of BKT transitions} \label{apdx:sinegordan}

Generally, the BKT transitions can be described by an effective sine-Gordon model \cite{Nomura_1995}
\begin{eqnarray}
\label{eq:sinegordon}
\mathcal{L}=\frac{1}{2 \pi K}(\nabla \phi)^{2}+\frac{y_{\phi}}{2 \pi \alpha^{2}} \cos \left( \sqrt{8} \phi \right),
\end{eqnarray}
where $\alpha$ is a ultraviolet cutoff. Writing $K = 1 + (1/2) y_0$, the RG equations under change of cutoff $\alpha \rightarrow e^{l} \alpha$ are
\begin{eqnarray}
\label{eq:sgrgeq}
\frac{d y_{0}(l)}{d l}=-y_{\phi}^{2}(l), \quad \frac{d y_{\phi}(l)}{d l}=-y_{\phi}(l) y_{0}(l)
\end{eqnarray}
Solving the RG equations, one obtains a line of stable fixed points for $y_{\phi} = 0, y_0 > 0$. The BKT critical lines are $y_0 = |y_{\phi}| > 0$, where the scaling dimension of $\cos \left( \sqrt{8} \phi \right)$ is $2$ (marginal). In the region $|y_{\phi}| < y_0$, the term $\cos \left( \sqrt{8} \phi \right)$ becomes irrelevant and all the points are renormalized onto the Gaussian fixed line, and are therefore massless. Outside this region, the field becomes relevant, all the points are renormalized away from the Gaussian fixed line, and are therefore massive. The BKT transition happens when a phase-transition-driving term moves the system across a BKT critical line. Near each BKT line in the massive phase, the energy gap scales as $\exp(-b/\sqrt{\delta t})$, where $\delta t$ is the distance to the BKT line \cite{Kosterlitz_1974}. If a system stays on the lines $y_0 = \pm y_{\phi}$, there is a phase transition from a massive phase into a BKT line across the SU(2) $\times$ SU(2) point at $y_0 = y_{\phi} = 0$, where the marginal fields disappear and the gap scales as $\sqrt{|y_0|}\exp(-b/|y_0|)$ \cite{PhysRevB.60.7850}. As the SG model becomes SU(2) symmetric on the BKT lines \cite{PhysRevD.12.1684, BANKS1976119}, systems with true BKT transitions would have an enhanced SU(2) symmetry at the phase-transition point, from which one can enumerate $7$ conditions for BKT transitions \cite{PhysRevB.100.094428}. Systems staying on $y_0 = \pm y_{\phi}$ lines should have a SU(2) symmetry for all parameter values. It has been shown that the O(2) model is equivalent to the SG model at $y_0 > 0, y_{\phi} > 0$ and have a true BKT transition \cite{PhysRevD.18.1916}. 

\section{Linear equations relating $\hat{U}^{\pm}$ and $\hat{S}^{\pm}$} \label{apdx:linearequations}

We discuss the solution for the linear system in Eq. \eqref{eq:auequalb}. The matrix elements $A_{ij}$ are exponentially large with $j$ for each $i < S-1$, so we expect the coefficients $u_q$ to be exponentially small with $q$. Figure \ref{fig:upmcoeffs} depicts the dependence of the magnitude of the coefficients $|u_q|$ on the index $q$ and confirms this expectation. Moreover, the absolute value of $u_q$ presents perfect exponential decay at first, then deviates up slightly. We emphasize that arbitrary precision arithmetic is required to obtain these results. We do a linear fit in the linear part and the slope becomes more negative as $S$ increases. From the inset Fig.~\ref{fig:upmcoeffs}(a), it is seen that the number of data points on the linear fits, $N_{fit}$, is proportional to the spin truncation. Then we plot the sign of $u_q$ as a function of $q$ in Fig.~\ref{fig:upmcoeffs}(b) for $S = 20$. The signs are initially consecutively positive for $q = 0,1, \ldots, 7$, and then oscillate between $+$ and $-$ for $q \ge 8$. This behavior is seen for all $S$. The number of consecutive positive signs before oscillation, $N_{plus}$, as a function of $S$ is plotted in Fig.~\ref{fig:upmcoeffs}(a), where we see that $N_{plus}$ is also proportional to $S$, and $N_{plus} \approx N_{fit}$. Finally, the slope of the linear fit in the main plot is a linear function of $\log{S}$, as shown in Fig.~\ref{fig:upmcoeffs}(c).

\end{document}